\documentclass[aps, prd, showpacs, nofootinbib, preprintnumbers,floatfix,longbibliography, reprint, superscriptaddress]{revtex4-1}
\usepackage[utf8]{inputenc}
\usepackage[colorlinks=true, linkcolor=blue, citecolor=teal]{hyperref}
\usepackage{bm}
\usepackage{latexsym,amssymb,amsmath,amsfonts}
\usepackage{graphicx}
\usepackage{longtable}
\usepackage{physics}
\usepackage{bm}
\usepackage{footnotebackref}
\usepackage{multirow}
\usepackage[usenames]{xcolor}
\usepackage{soul}
\usepackage{makecell}
\usepackage{graphicx}
\usepackage{amssymb}
\usepackage{amsmath}
\usepackage{longtable}
\usepackage{rotating}
\usepackage{color}
\usepackage{fancyhdr}
\usepackage{ifthen}
\usepackage{slashed}
\usepackage{xspace}
\usepackage{tabularx}
\usepackage{float}

\begin{document}

\title{Angular power spectra of anisotropic stochastic gravitational wave background: developing statistical methods and analyzing data from ground-based detectors}

\author{Deepali Agarwal}
\email{deepali@iucaa.in}
\affiliation{Inter-University Centre for Astronomy and Astrophysics (IUCAA), Pune 411007, India}

\author{Jishnu Suresh}
\email{jishnu.suresh@uclouvain.be}
\affiliation{Centre for Cosmology, Particle Physics and Phenomenology (CP3), Universit\'e catholique de Louvain, Louvain-la-Neuve, B-1348, Belgium}

\author{Sanjit Mitra}
\email{sanjit@iucaa.in}
\affiliation{Inter-University Centre for Astronomy and Astrophysics (IUCAA), Pune 411007, India}

\author{Anirban Ain}
\email{anirban.ain@pi.infn.it}
\affiliation{Istituto Nazionale di Fisica Nucleare (INFN) sezione Pisa, 56126 Pisa, Italy }
%


\begin{abstract}

Unresolved sources of gravitational waves can create a stochastic gravitational wave background (SGWB) which may have intrinsic or extrinsic anisotropies. The angular power spectrum is a well-suited estimator for characterizing diffuse anisotropic distributions in the sky. Here we estimate the first model-independent all-sky all-frequency SGWB angular power spectra in the 20-1726 Hz frequency range from the third observing run (O3) of the Advanced LIGO and Advanced Virgo detectors. 
We develop a method to use the spectrum's signal-to-noise ratio as the detection statistic and show that the shape of the distribution of the statistic obtained from the data agrees with the analytical model with a modified value of the parameter. Since we find the data to be consistent with noise, $95\%$ confidence Bayesian upper limits are set on the angular power spectra, ranging from $C_\ell^{1/2}\leq(3.0\times10^{-9}-0.73)~\text{sr}^{-1}$. We also introduce a method to combine the narrow band angular power spectra to obtain estimators for broadband SGWB. These results can directly constrain theoretical models that predict the SGWB angular power spectra and for estimating or constraining the corresponding parameters. In addition, the results and the techniques introduced in this work can be useful for performing correlation-based searches, for instance, with electromagnetic observations.
\end{abstract}


\maketitle


\section{Introduction}
While more than 90 compact binary coalescences are being cataloged~\cite{GWTC3,OGC4}, the search for other kinds of gravitational wave (GW) sources is continuing with great enthusiasm. The stochastic gravitational wave background is one of the potential sources for detection in the coming years with the network of ground-based GW observatories. The observed rates of compact binary mergers suggest that the dominant contribution to this background will likely be from the superposition of signals from such mergers throughout the Universe~\cite{Regimbau_2011,Regimbau:2022mdu}. Along with the mergers of black holes and neutron stars, many different sources will contribute to the astrophysical stochastic gravitational wave background (SGWB), e.g., inspiraling binaries, supernova explosions, and spinning neutron stars. 

It has been shown that the energy flux from all astrophysical sources (resolved and unresolved) is not constant across the sky and depends on the direction of observation~\cite{Jenkins:2018lvb,2014PhRvD..89h4076M,PhysRevLett.120.231101, jenkins_sakell_cbc_anisotropy,2012PhRvD..86j4007R,2013PhRvD..87d2002W,2013PhRvD..87f3004L,PhysRevD.96.103019,Capurri_2021,PhysRevD.101.103513}. Several factors result in such anisotropies: the specific distribution of astrophysical SGWB sources, anisotropy accumulated along the line-of-sight, gravitational lensing, redshift-space distortions, and dipole anisotropy induced by the observer's peculiar velocity. The stochastic directional radiometer search can provide information on the angular content of the SGWB in the form of a skymap (pixel basis) or its spherical harmonic (SpH) coefficients. The anisotropic SGWB search is expected to be powerful in identifying and characterizing the individual contributors to the total stochastic background~\cite{Cusin_2022,PhysRevD.106.082005}. 

The anisotropic SGWB searches estimate the GW energy density (usually in the units of critical density needed for a flat universe) per unit frequency while retaining the directional information $\Omega (f, \Theta)$. The standard, so far, has been to perform the search and present results for cases where $\Omega (f, \Theta)$ is assumed to follow a power law frequency variation (a model-dependent approach).
Sky maps have been produced for all the observing runs of the advanced ground-based interferometric detectors~\cite{O3directional}. While these maps are created in pixel and SpH bases to probe localized and diffuse sources, employing cross-correlation-based algorithms, the underlying algebra and numerical implementation remain different~\cite{Ballmer_2006,mitra07,eric_sph}. Consequently, there was a need to produce sky maps on both bases. Recently, it has been shown that these manifestly redundant methods could be unified into a single analysis that can probe very different scales and demonstrate unification using real data~\cite{pystoch_sph,liting}. 

Previous anisotropic SGWB searches were either targeted (pointing in one direction) and narrow band (considering many different frequencies) or all-sky (looking in all directions) but broadband (averaging over all frequencies). Hence it had limited prospects of detecting an unknown narrow band anisotropic SGWB. To mitigate this, the directional search must be conducted separately across all narrow frequency bins. This demands a lot of computational power. However, exploiting a temporal symmetry in the map-making algebra, the conventional searches can be made a few hundred times faster through the data folding mechanism~\cite{folding}. Recently, together with a {\tt PYTHON}-based map-making pipeline, {\tt PyStoch}, the traditional analysis was made hundreds of times faster and opened up the possibility of performing an extremely efficient search looking in all the directions and at all frequencies [all-sky all-frequency (ASAF)]~\cite{pystoch,pystoch_sph}. Recently this method was implemented, for the first time, on the data from LIGO-Virgo-KAGRA's first three observational runs~\cite {asaf_lvk}. Other efforts have been towards the same goal~\cite{Renzini_aniso,liting}.

The ASAF search targets SGWB from unknown signals in narrow frequency bins (producing sky maps at each frequency bin considered in the analysis), providing a sensitive tool for discovering any persistent source which does not conform to the assumptions made by template-based (matched filtering type) searches. One can then find the pixels in the maps containing statistically significant outliers, which can, for example, be followed up with a more sensitive search. While one can also derive the SpH coefficients of the narrow band maps, the method to find candidates for diffuse sources (for which the SpH basis is more suitable) needed to be carried out. Unlike compact binary mergers, the many narrow band sources, e.g., extragalactic young and millisecond pulsars, may undergo negligible frequency evolution over observing timescales. This could lead to a frequency-dependent angular power spectrum due to a distinct source population. This requires coming up with an appropriate detection statistic, along with its expected probability distribution. This paper presents the angular power spectra of the anisotropic SGWB at every frequency bin using data from the third observing run of Advanced LIGO~\cite{advLIGO} and Advanced Virgo~\cite{advVirgo} detectors. We assign significance to the data using the angular power spectra statistic after obtaining the distribution of its noise background. A narrow band estimator like this is manifestly independent of the frequency spectral model. These estimates will complement the current efforts in understanding the anisotropies associated with the SGWB. 

The paper is organized as follows: In Sec.~\ref{sec:method}, we discuss the approach to map the anisotropy of SGWB in a SpH basis and compute the angular power spectra for narrow band background. Then, the details of the analysis and results are presented in Sec.~\ref{sec:results}. Finally, the article is summarized in Sec.~\ref{sec:conclusion} along with future prospects of the search. 



\section{Method}\label{sec:method}
The SGWB is characterized by its dimensionless spectrum defined in terms of a frequency and direction-dependent form as
\begin{equation}
    \Omega_{\mathrm{GW}} (f, \Theta) = \frac{f}{\rho_c} \frac{\dd \rho_{\mathrm{GW}}(f, \Theta)}{\dd f}\,,
\end{equation}
where $\dd \rho_{\mathrm{GW}}$ is the GW energy density contained in the $f$ and $f+\dd f$ frequency range, $\Theta$ represents the directions on the sky, and $\rho_c$ is the critical energy density needed to close the Universe. 
In the past broadband searches for SGWB anisotropy~\cite{O3directional}, it was assumed that one could factorize the above spectrum into a frequency-dependent part $H(f)$ and a direction-dependent part $\mathcal{P}(\Theta)$. The frequency dependence of the spectrum through $H(f)$ is typically assumed to take a power-law form consistent with the most canonical models for SGWB. As mentioned in the recent all-sky all-frequency radiometer search~\cite{asaf_lvk}, one can unfetter this assumption by performing a narrow band search, as it is inherently model independent. Following the past conventions~\cite{romanoreview}, one can rewrite the above equation as
\begin{equation}\label{eq:Omega_GW}
    \Omega_{\rm GW}(f,\Theta)  =  \frac{2\pi^2}{3H^2_0}f^3 \, \mathcal{P}(f,\Theta)\ ,
\end{equation}
where $H_0$ is the Hubble constant taken to be $H_0=67.8$ kms$^{-1}$Mpc$^{-1}$~\cite{HubblePlanck} and $\mathcal{P}(f,\Theta)$ records the angular variation of the one-sided power spectral density of the SGWB.
It is shown in the literature that, for a diffuse background, the SpH basis is a better choice for the searches. A method for obtaining maximum likelihood (ML) sky maps in SpH basis was developed and tested in \citet{eric_sph} and~\citet{pystoch_sph}. Using the SpH basis, it is possible to expand and map the direction and frequency-dependent $\mathcal{P}(f,\Theta)$ component in terms of spherical harmonics $Y_{\ell m}$ as
\begin{equation}
\mathcal{P}(f,\Theta) \ = \ \sum_{\ell=0}^{\ell_\textrm{max}}\sum_{m=-\ell}^{\ell}\mathcal{P}_{\ell m}(f)\, Y_{\ell m}(\Theta) \, ,
\label{POmega:spherical_basis}
\end{equation}
where $\mathcal{P}_{\ell m}(f)$ are the SpH coefficients. Ideally, the value of the highest $\ell$ mode should be $\ell_\textrm{max}=\infty$, but in practice, it is limited by the angular resolution achieved by the detector network as
\begin{equation}
\ell_\textrm{max}=\frac{2\pi\,d\,f}{c}\,,
\end{equation}
where $d$ is the baseline length and $c$ is the speed of the light~\cite{eric_sph,PhysRevLett.118.121102,Erik_Floden}. 

For a statistically isotropic Gaussian background\footnote{The Gaussianity of the background can be achieved by satisfying certain criterion at any angular resolution by increasing the observation time~\cite{Jenkins:2018lvb,jenkins_sakell_cbc_anisotropy}.}, the mean, $\bar{\mathcal{P}}$, defined as~\cite{jenkins_thesis}
\begin{equation}
    \langle\mathcal{P}_{\ell m}(f)\rangle_U =\sqrt{4\pi}\,\bar{\mathcal{P}}(f)\,\delta_{\ell 0}\,,
\end{equation}
and the covariance is defined as
\begin{equation}
    \textrm{Cov}[\mathcal{P}_{\ell m}(f)\mathcal{P}_{\ell' m'}(f)]_U=\left(\frac{2\pi^2f^3}{3H^2_0}\right)^{-2}\,C_\ell(f)\,\delta_{\ell\ell'}\,\delta_{mm'}\,,
\end{equation}
provides a complete statistical description of the anisotropic sky.
Otherwise, though the statistical description provided by $\mathcal{C}_\ell(f)$ is incomplete, the information can be useful to measure power at different angular scales and identify deviations from noise, which can be useful for detecting anisotropic distributions. The average $\langle\cdot\rangle_U$ is evaluated over an ensemble of the Universe. $C_\ell(f)$ is the angular power spectrum of the sky. We define the observed angular power spectrum using the SpH coefficients as~\cite{eric_sph}
\begin{equation}
    \tilde{C}_\ell(f) =\left(\frac{2\pi^2f^3}{3H^2_0}\right)^{2}\, \frac{1}{2\ell+1}\sum_{m=-\ell}^\ell |\mathcal{P}_{\ell m}(f)|^2\,,
\end{equation}
which is an unbiased estimator of the true angular power spectrum, $C_\ell(f)$. $C_\ell(f)$ has units of sr$^{-2}$.

We are interested in estimating the narrow band angular power spectrum in this paper. For that, we start with the cross-spectral density of the data~\cite{pystoch}, $s_{\mathcal{I}_{1,2}}(t)$, from a pair of GW detectors ($\mathcal{I}_{1}$ and $\mathcal{I}_2$),
\begin{equation}
     \mathcal{C}^\mathcal{I}(t;f) = \frac{2}{\tau}\,\tilde{s}_{\mathcal{I}_{1}}^*(t;f) \,\tilde{s}_{\mathcal{I}_{2}}(t;f)  \,,
\end{equation}
where $\tilde{s}_{\mathcal{I}_{1,2}}(t;f)$ are the short-term Fourier transform of $s_{\mathcal{I}_{1,2}}(t)$ of a segment centered at time $t$ with duration $\tau$ and $f$ are (positive and negative) frequencies.
The expectation of the cross-spectral density is given by~\cite{pystoch_sph}
\begin{equation}
     \langle \mathcal{C}^\mathcal{I}(t;f) \rangle_N = \sum_{lm} \gamma_{ft,\ell m}^{\mathcal{I}} \, \mathcal{P}_{\ell m}(f) \,,
\end{equation}
where the average $\langle\cdot\rangle_N$ is evaluated over an ensemble of the noise realizations and $\gamma_{ft,\ell m}^{\mathcal{I}}$ is the generalized overlap reduction function~\cite{christ92,ORF_Finn}, which accounts for the mismatch between the response functions of the detectors and the delay in signal arrival times, defined as
\begin{equation}
\gamma_{ft,\ell m} ^{\mathcal{I}} = \int_{S^2} d \Theta\, \gamma_{ft,\Theta}^{\mathcal{I}} \, Y _{\ell m}(\Theta) \,.
\label{eq_ORF_spherical}
\end{equation}

To measure the anisotropy $\mathcal{P}_{\ell m}(f)$, the radiometer algorithm uses the ML estimator as the statistic~\cite{eric_sph}. The model-independent SpH coefficients that maximize the likelihood function are given by~\cite{eric_sph,asaf_lvk}
\begin{equation}
    \hat{\mathcal{P}}_{\ell m}(f) = \Gamma_{\ell m,\ell'm'}^{-1} (f)\, X_{\ell'm'}(f)\,,
    \label{eq:clean_plm}
\end{equation}
where,
\begin{equation}
X_{\ell m}(f) \ = \tau\,\Delta f\, \sum_{\mathcal{I}t}  \frac{ \gamma^{\mathcal{I}*}_{ft,\ell m}\,C^{\mathcal{I}} (t;f)} {P_{\mathcal{I}_1}(t;f) P_{\mathcal{I}_2}(t;f)}  \,, 
\label{eq:Dirty_map}
\end{equation}
and
\begin{equation}
\Gamma_{\ell m,\ell'm'} (f) =  \tau\,\Delta f\, \sum_{\mathcal{I}t} \frac{\gamma^{\mathcal{I}*}_{ft,\ell m} \, \gamma^{\mathcal{I}}_{ft,\ell'm'}}{P_{\mathcal{I}_1}(t;f) \, P_{\mathcal{I}_2}(t;f)}\, .
\label{eq_fisher_sph}
\end{equation}
The ASAF SpH dirty map shown in Eq.~(\ref{eq:Dirty_map}) denotes the SGWB anisotropic sky observed through the antenna response pattern of the detector pair used to form the baseline $\mathcal{I}$. In this equation, $P_{\mathcal{I}_{1,2}}(t;f)$ denotes the one-sided noise power spectra of the detector output for the time segment $t$. The covariance matrix of the dirty map in the weak signal limit is given in Eq.~(\ref{eq_fisher_sph}), and it is often called the Fisher information matrix.

The ML estimators of the angular and frequency distribution of the SGWB power, given in Eq.~(\ref{eq:clean_plm}), are usually referred to as the clean maps since they are estimators of the actual GW sky, obtained by deconvolving the detector responses from the dirty maps. As evident from the equation, the deconvolution demands the computation of the inverse of the Fisher information matrix to obtain the clean map. However, in practice, the Fisher matrix is poorly conditioned due to the diffraction limit and blind directions of the detector or detector network. Consequently, one must regularize the Fisher matrix before the inversion. Even though many techniques~\cite{eric_sph,romanoreview,sambit} can be used to regularize the matrix\footnote{Regularization techniques include cutting off the eigenvalues of the Fisher matrix at some specific $\ell_{\mathrm{max}}$ values, getting a matrix with reduced rank by modifying the eigenvalues, and only considering the diagonal components of the Fisher matrix ignoring all off-diagonal correlations.}, in this work, we use the singular value decomposition (SVD) method, which has been proposed and tested for the SGWB searches in \citet{eric_sph}. Note that this regularization introduces a bias in our estimators. The bias can be estimated if the power distribution is known, which is not the case for most astrophysical scenarios. The Fisher matrix at every frequency bin is Hermitian, which is evident from its definition, so its SVD takes the form,
\begin{equation}
\mathbf{\Gamma} (f) = \mathbf{U} \mathbf{S} \mathbf{V}^\dagger \,,
\label{fisher_svd}
\end{equation}
where $\mathbf{U}$ and $\mathbf{V}$ are unitary matrices, and $\mathbf{S}$ is a diagonal matrix whose nonzero elements are the real and positive eigenvalues of the Fisher matrix, arranged in descending order. To condition the matrix, a threshold $S_{\rm min}$ is chosen. The choice is made by considering the proper trade-off between the quality of the deconvolution and the increase in numerical noise from less sensitive modes. Any values below this cutoff are considered too small, and we replace them with infinity. This is to prevent the inverted noise from the corresponding insensitive modes. Now, one can write the inverse of regularized Fisher matrix, which is obtained using the modified matrix $\mathbf{S}_R$  as
\begin{equation}
\mathbf{\Gamma}_R^{-1} (f) = \mathbf{V} \mathbf{S}_R^{-1} \mathbf{U}^\dagger\,.
\end{equation}
By multiplying the inverted-regularized Fisher matrix with the dirty map, one can obtain the estimators of the SpH coefficients:
\begin{equation}\label{eq:clean_plm_reg}
\mathcal{\hat{P}}_{\ell m} (f) = (\mathbf{\Gamma}_R^{-1})_{\ell m,\ell'm'}(f)\,X_{\ell'm'} (f),
\end{equation} 
and their uncertainty can be written as~\cite{eric_sph}
\begin{eqnarray}\label{eq:clean_sig_plm}
\sigma_{\ell m} (f) &=& \sqrt{\mathrm{Var}[\mathcal{\hat{P}}_{\ell m} (f)]}\\
&=& \sqrt{\left[\mathbf{\Gamma}_R^{-1} (f) \, \mathbf{\Gamma}(f) \, \mathbf{\Gamma}_R^{-1} (f) \right]_{\ell m,\ell m} }\,.
\end{eqnarray}

One can use the above clean map in the SpH basis to construct the unbiased\footnote{The expected bias due to noise covariance is subtracted; however, some bias remains due to regularized deconvolution. As mentioned before, this could not be accounted for if the expected source angular power spectra were \textit{a priori} unknown.} estimator of the narrow band angular power spectrum~\cite{eric_sph},
\begin{eqnarray}
\label{ang_power_spectra}
\hat{C_\ell} (f) =\left(\frac{2\pi^2f^3}{3H^2_0}\right)^{2}\, && \frac{1}{2 \ell + 1} \sum_{m} \bigg[|\hat{\mathcal P}_{\ell m} (f)|^2 \nonumber \\
&&- \left[\mathbf{\Gamma}_R^{-1} \mathbf{\Gamma} \mathbf{\Gamma}_R^{-1} \right]_{\ell m,\ell m} (f) \bigg] .
\end{eqnarray}
Similar to the clean map, one can write the covariance matrix of the above angular power spectrum estimator as
\begin{eqnarray}\label{eq:C_l_Cov}
    \mathbf{\Sigma}(f) \equiv \Sigma_{\ell\ell'}(f) 
    =&&\left(\frac{2\pi^2f^3}{3H^2_0}\right)^{4}\,\frac{2}{(2\ell+1)(2\ell'+1)}\nonumber\\
    &&\sum_{m,m'} |\mathbf{\Gamma}_R^{-1} \, \mathbf{\Gamma} \, \mathbf{\Gamma}_R^{-1}  |^2_{\ell m,\ell'm'}(f) \,, \nonumber\\
\end{eqnarray}
whose diagonal elements are the measures of the standard deviations of the estimators of the angular power spectrum, i.e., $\sigma_{C_\ell}(f)=\sqrt{\Sigma_{\ell\ell}(f)}$ (see Appendix~\ref{sec:covMatDerivation} for the derivation).
Given that we have both the angular power spectra and the uncertainty associated with each measurement, we can define the signal-to-noise ratio (SNR) as
\begin{equation}\label{eq:ang_power_spectra_snr}
    \rho_\ell (f)= \frac{\hat{C_\ell} \, (f)}{\sigma_{C_\ell} \, (f) } \, .
\end{equation}

The exact analytic expression for the probability density function (PDF) of SNR, $\rho_\ell (f)$, is nontrivial to derive as, in practice, the noise corresponding to different $m$ modes have different variances, and the correlation between different $\ell$ modes introduces further complexities. Nevertheless, here we assume that the clean SpH coefficients are uncorrelated and the noise for each $m$ mode for a given $\ell$ mode is white, i.e.,
\begin{equation}\label{eq:snrpdfassum}
    \left[ \mathbf{\Gamma}_R^{-1} \, \mathbf{\Gamma} \, \mathbf{\Gamma}_R^{-1} \right]_{\ell m,\ell'm'} (f) = \sigma^2_\ell(f)\,\delta_{\ell \ell'}\delta_{mm'}\,.
\end{equation}
Then the SNR can be written as
\begin{equation}
    \rho_\ell (f)=\frac{1}{\sqrt{2(2\ell+1)}}\sum_{m}\left[|\rho_{\ell m}(f)|^2-1\right]\,,
\end{equation}
where $\rho_{\ell m}$ is SNR of clean SpH coefficients $\hat{\mathcal P}_{\ell m} (f)$. Following the central limit theorem, it can be assumed to follow a normal distribution in the noise-only case. Then the PDF for SNR  $\rho_\ell (f)$ is a chi-squared distribution with degrees-of-freedom (DOF) of $k=2\ell+1$ as
\begin{equation}
P(y=\rho_\ell (f))\,dy=\sqrt{2k}\,\chi^2_{k}(y\sqrt{2k}+k)\,dy\,.
\end{equation}

In the next section, we show that the shape of this analytical model matches the numerically obtained distribution with a modified DOF. The difference between modified DOF $k_{\textrm{eff}}$ and true DOF $k$ indicates the deviation from our assumption of uncorrelated and white noise modes. (See Appendix~\ref{sec:apprxPDF} for the details.)

In this work, $\rho_\ell (f)$ is used as the detection statistic to assign significance to the data, and in case of no detection, we set constraints on narrow band angular power spectra $C_\ell(f)$. 



\section{Implementation and Analysis}\label{sec:results}
To perform the search, we analyze time-series data from the third (O3) observing run of the Advanced LIGO~\cite{advLIGO} Hanford (H) and Livingston (L) detectors and the Advanced Virgo (V) detector. We first apply time and frequency domain data quality cuts, identically as was done in~\citet{asaf_lvk}. The cleaned data is then folded to one sidereal day, which utilizes the temporal symmetry in the map-making algorithm~\cite{folding}. Folding reduced the computation cost by a factor equal to the total number of sidereal days of coincident quality data. This reduction was critical for performing this analysis. The folded datasets for the first three observing runs of Advanced LIGO and Advanced Virgo detectors are publicly available~\cite{ligo_scientific_collaboration_virgo_coll_2022_6326656}. Next, the narrow band (1 Hz bandwidth) dirty maps and the Fisher information matrices in the SpH basis are computed for the frequency range 20-1726 Hz using the {\tt PyStoch} code~\cite{pystoch,pystoch_sph}. This pipeline brought additional computational advantages and the power of {\tt HEALPix}~\cite{HEALPix,Zonca}, which was also crucial for this analysis. 

\begin{figure}[h]
    \centering
    \includegraphics[width=\columnwidth]
    {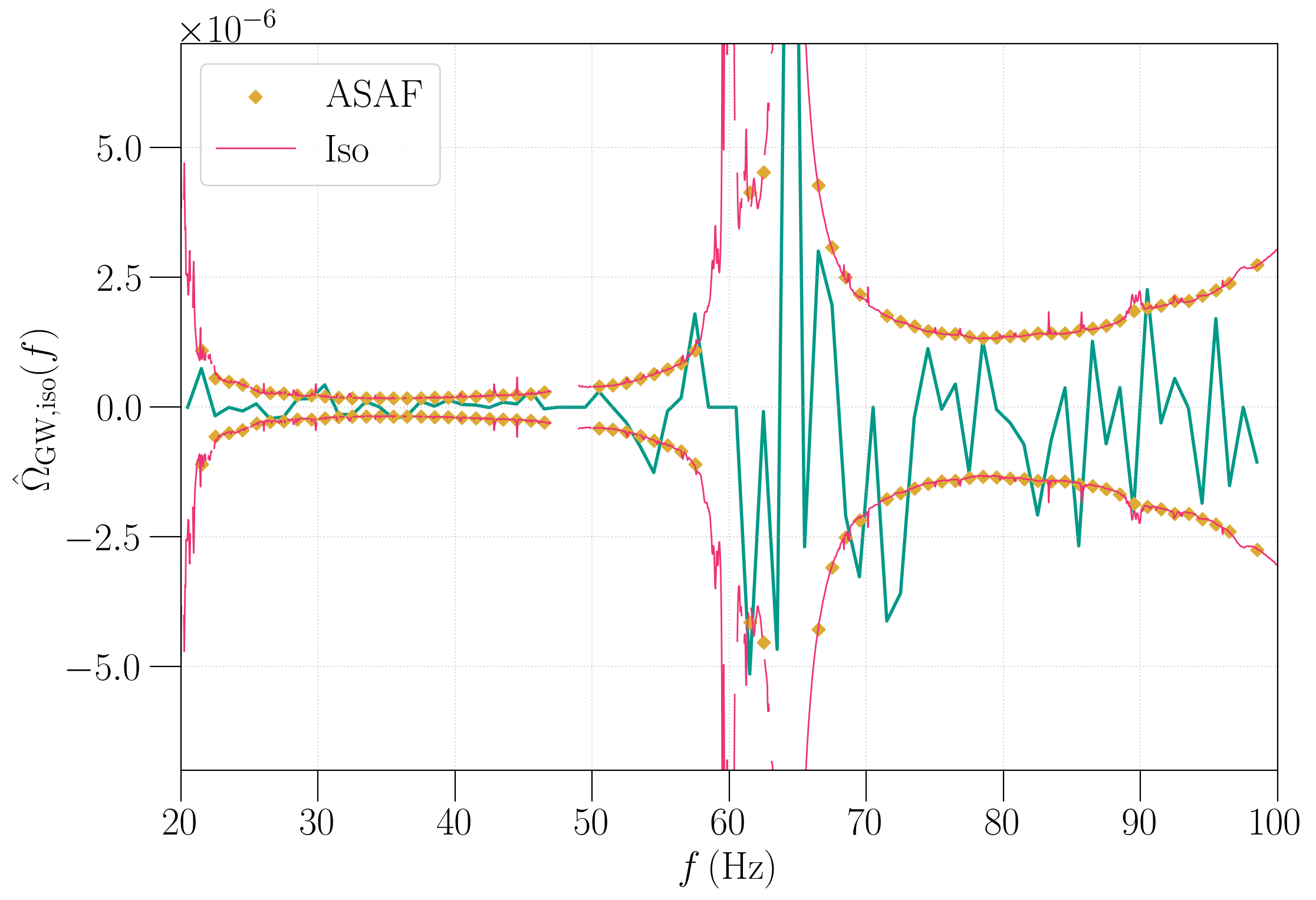}
    \caption{The ML estimator of monopole (ASAF-SpH; $\ell_\mathrm{max}=0$) as a function of frequency (green line) and its uncertainty compared with the error bars (orange scatter) of the estimator obtained from O3 isotropic analysis (red line)~\cite{O3iso} using the HL baseline.}
    \label{fig:monopole}
\end{figure}

\subsection{Monopole}

\begin{table}[b]
    \centering
    \begin{tabular}{c|c|c}
        \hline\hline
        \multicolumn{3}{c}{Broadband Results : HL baseline } \\
        \hline
       \hline
       \multicolumn{1}{c|}{$\alpha$}& \multicolumn{2}{c}{$\hat{\Omega}_\mathrm{GW,iso} (\times 10^{-9}$)}\\
        \hline
        \multicolumn{1}{c|}{}& ASAF-SpH & ISO~\cite{O3iso}\\
         \hline
          0 &  2.4 $\pm$ 8.6  & -2.1 $\pm$ 8.2  \\
         \hline
         2/3 &   0.55 $\pm$ 6.5  & -3.4 $\pm$ 6.1\\
         \hline
          3 & -0.56 $\pm$ 1.0 & -1.3 $\pm$ 0.9\\
         \hline\hline
    \end{tabular}
    \caption{Broadband isotropic search results derived from the ASAF-SpH monopole term (with the stricter notching) using the O3 HL dataset. Results from the previous LVK O3 analyses are also added to the table for a direct comparison. The three spectral indices used in the search are denoted by $\alpha$.}
    \label{tab:ISO_Results}
\end{table}


\begin{figure*}
    \centering
    \includegraphics[width=\columnwidth]
    {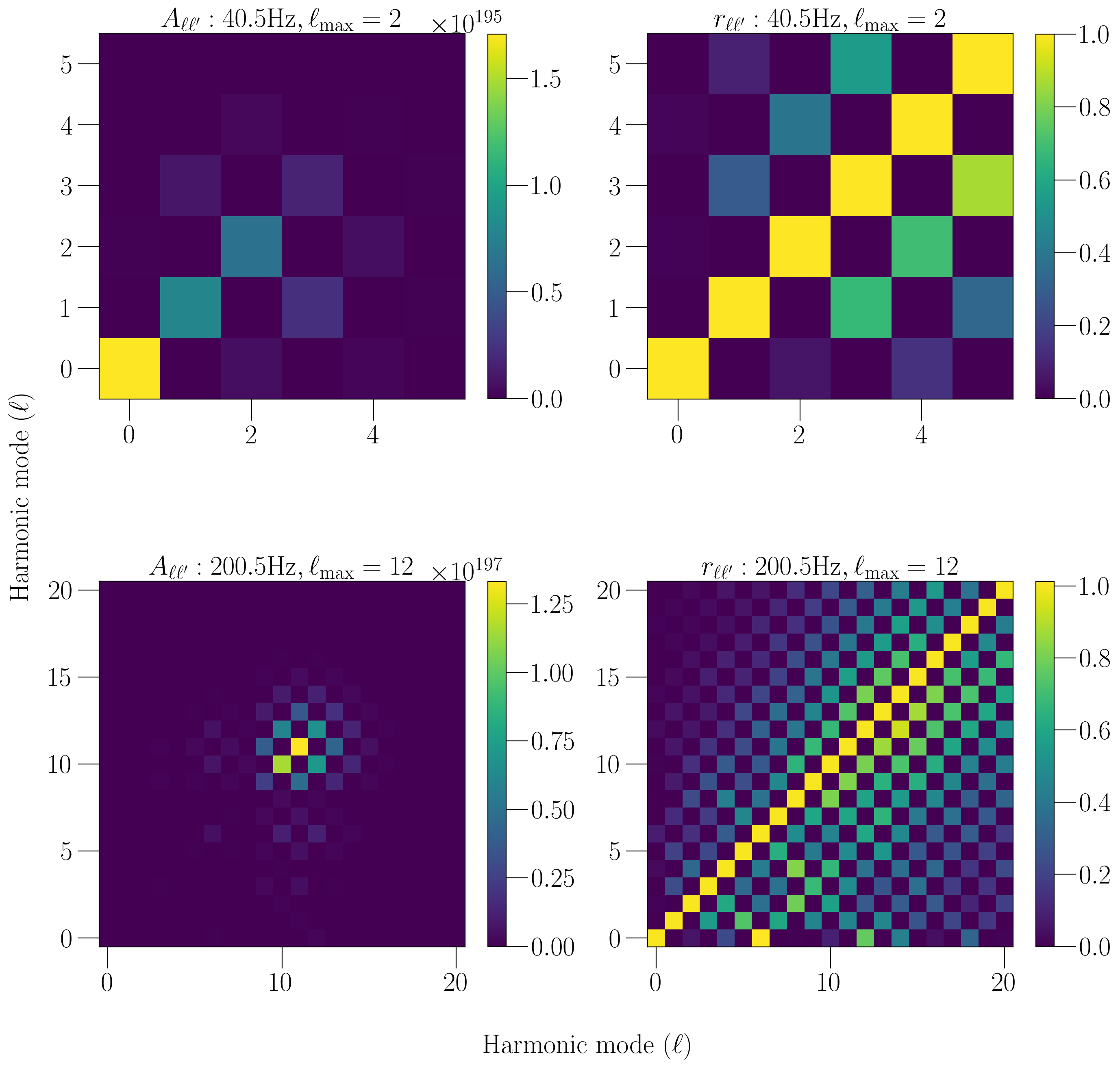} ~~~~
    \includegraphics[width=\columnwidth]
    {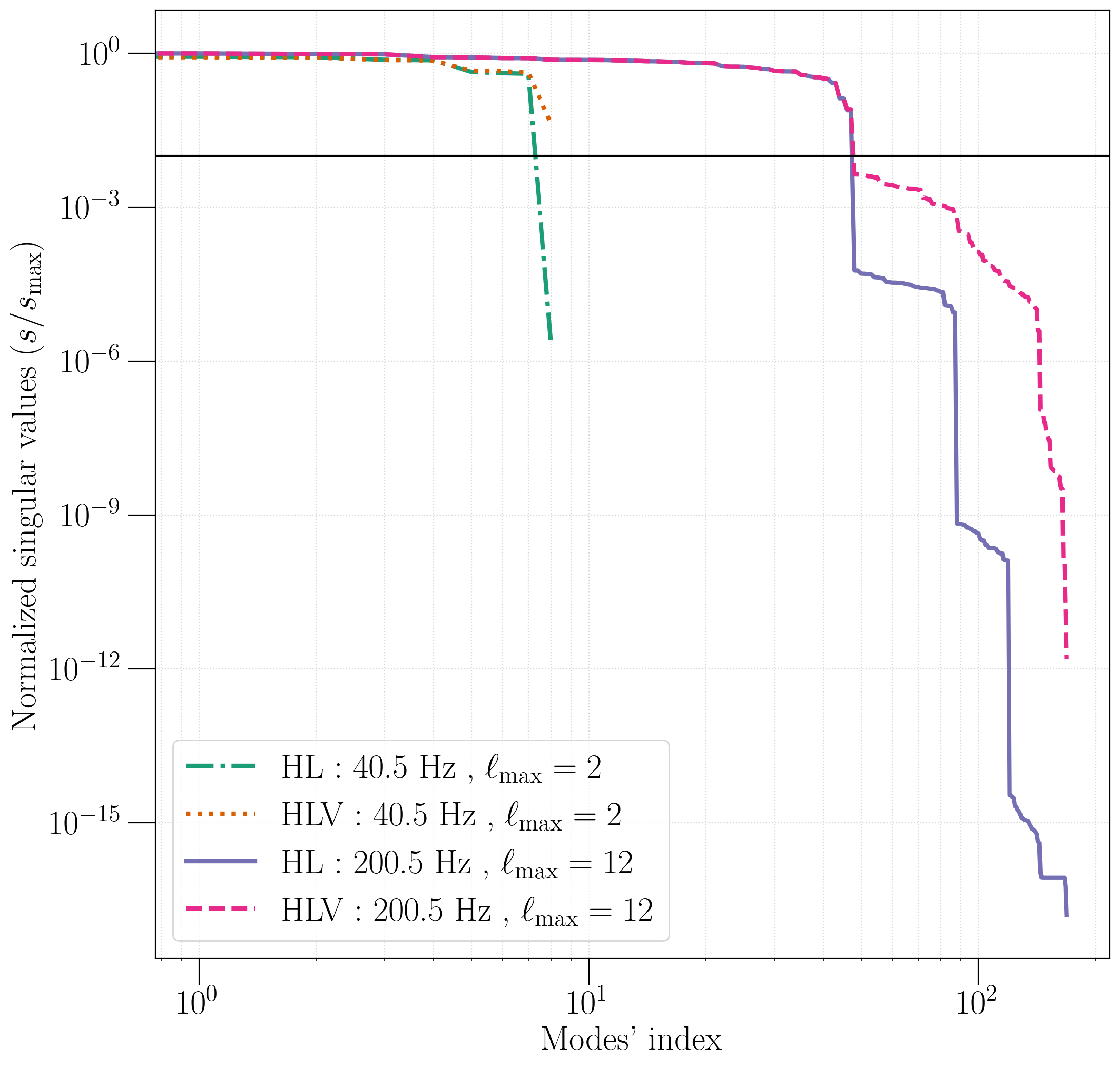} 
    \caption{The matrix plots on the left present the measure of the detector response, $\mathbf{A}$, and the expected correlation between the SpH modes measured using HL baseline at frequencies 40.5 and 200.5 Hz, respectively. The correlation is quantified by the ratio $r_{\ell\ell'}=\frac{A_{\ell\ell'}}{\sqrt{A_{\ell\ell}\,A_{\ell'\ell'}}}$. The correlation and detector response is found to be a function of frequency. The right plot depicts the singular value spectrum for the Fisher matrix, $\mathbf{\Gamma}$, obtained with the HL and HLV baselines. It shows that including the Virgo baseline slightly improves the ill-conditioned nature of the matrix.}
    \label{fig:covMatDemo}
\end{figure*}

For an isotropic background, the dimensionless energy density parameter, $\Omega_{\mathrm{GW}} (f, \Theta)=  \Omega_{\mathrm{GW}} (f)$ is related to the monopole coefficient $\mathcal{P}_{00}(f)$ as~\cite{eric_sph}
\begin{equation}\label{eq:monopole}
    \Omega_{\mathrm{GW}} (f) = \frac{2\pi^2}{3H_0^2}\,f^3\,\sqrt{4\pi}\,\mathcal{P}_{00}(f)\,,
\end{equation}
since the contribution from all higher multipoles is zero when integrated over the whole sky. We present a comparison of the estimator obtained in this analysis (ASAF-SpH) and the results from narrow band isotropic searches\footnote{The error bar $\sigma_{00} (f)$ is multiplied with the normalization constant given in Eq.~(\ref{eq:monopole}) and the square root of the number of unnotched bins in each frequency bin to account for the difference in frequency resolution used in analyses.} in Fig.~\ref{fig:monopole}~\cite{O3iso}. The results are in good agreement. The ML estimator and its error bar for a broadband background can be obtained by combining the dirty maps and the Fisher matrix as
 \begin{equation}
X_{\ell m} \ = \ \sum_{f}  H(f)\, X_{\ell m} (f)\,, 
\label{eq:Dirty_map_BBR}
\end{equation}
and
\begin{equation}
\Gamma_{\ell m,\ell'm'} =  \sum_{f}H^{2}(f)\, \Gamma_{\ell m,\ell'm'} (f) \, ,
\label{eq_fisher_sph_BBR}
\end{equation}
and using Eqs. (\ref{eq:clean_plm}), (\ref{eq:clean_sig_plm}), (\ref{eq:monopole}) with $\ell_{\text{max}}=0$ for an isotropic background. The broadband background is expected to follow a power law spectrum defined as $H(f)=H(f_{\text{ref}})(f/f_{\text{ref}})^{\alpha-3}$, where $\alpha$ is the spectral index and $f_{\text{ref}}$ is the most sensitive frequency of the observing band which is chosen to be 25 Hz. A comparison is shown in Table.~\ref{tab:ISO_Results}. The match between these (model-dependent) results proves the consistency of converting pixel basis maps to SpH basis maps.

\subsection{Narrow band higher multipoles}

\begin{figure*}
    \centering
    \includegraphics[width=\textwidth]
    {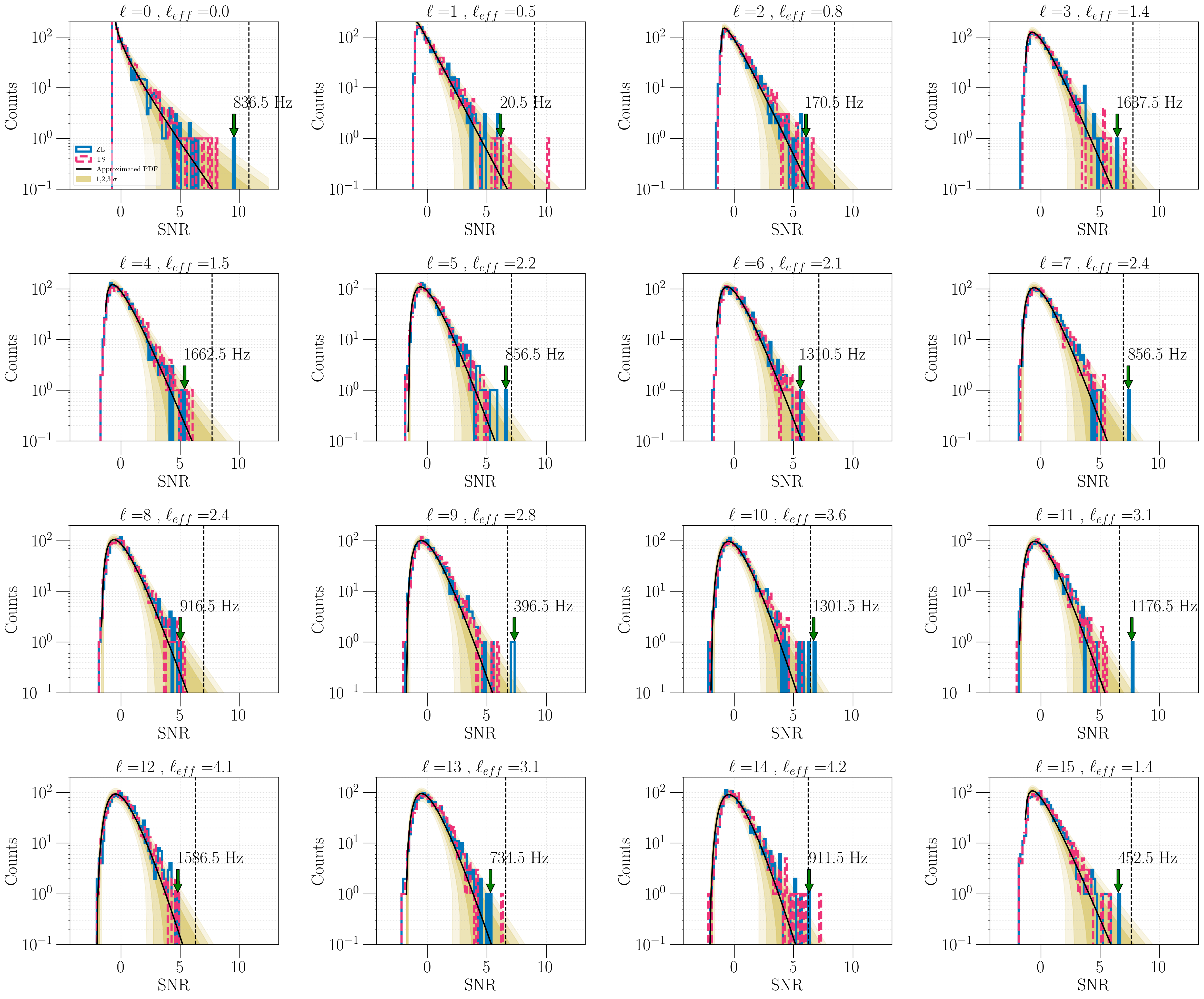}
    \caption{Histogram of SNRs obtained using the zero-lag (blue) and time-shifted (pink) data from the third observing run of Advanced LIGO and Advanced Virgo (HLV) detectors. Here, for each harmonic mode, the frequency samples are treated as independent samples for the statistic. The solid black line represents the approximate PDF with effective DOF $k_{\textrm{eff}}=2\,\ell_{\textrm{eff}}+1$ best fitted to the samples obtained by the time-shifted run. The zero-lag and time-shifted data is broadly consistent with the approximated distribution of SNRs within three-sigma (yellow color) Poisson error bars for lower harmonic modes but deviate at higher modes. The frequencies for maximum SNR in the zero-lag run are indicated in the histogram. The black dashed line depicts the SNR threshold for the global p value 0.05 given approximated PDF.}
    \label{fig:histogram}
\end{figure*}

\begin{figure*}
    \centering
    \includegraphics[width=\columnwidth]
    {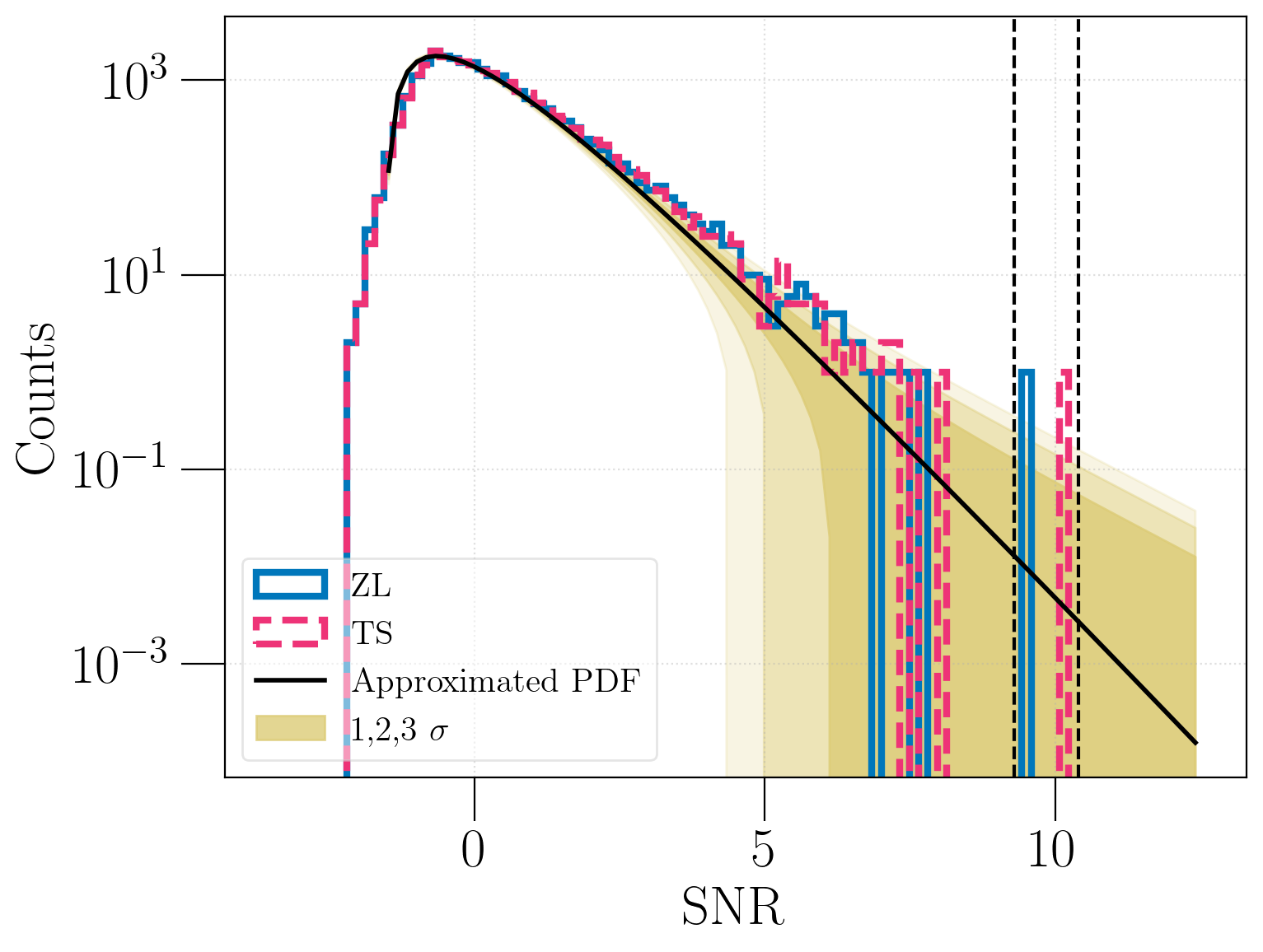}~~~
    \includegraphics[width=\columnwidth]
    {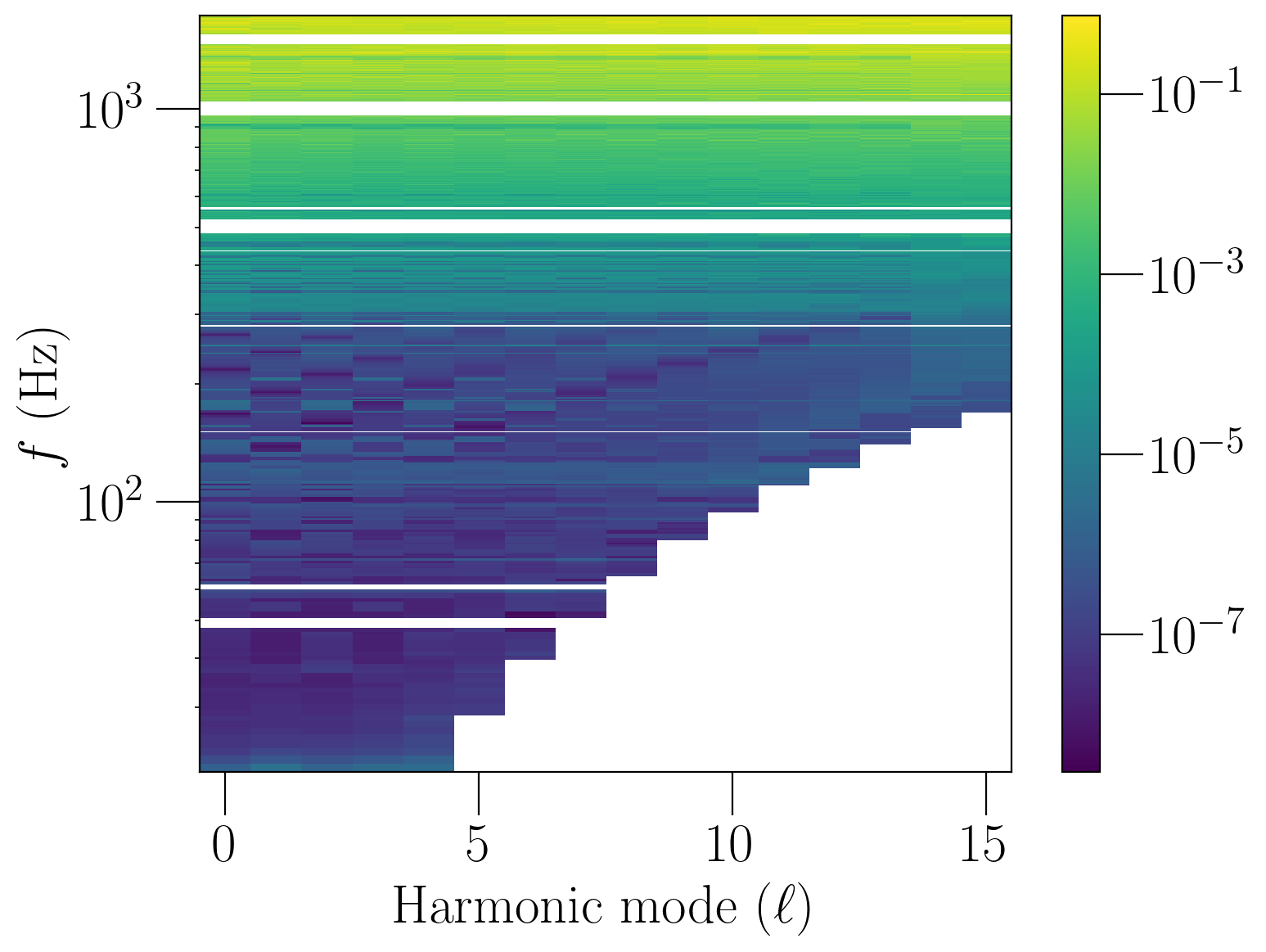}
   \caption{Left: joint histogram of the SNR samples obtained using the zero-lag (ZL; blue) and time-shifted (TS; pink) data from the third observing run of Advanced LIGO and Advanced Virgo (HLV) detectors. The solid black line represents a fitted approximated PDF with $\ell_\mathrm{eff}=1.82$. The vertical dashed lines show the SNR threshold for the global p value 5\% and 1\%, respectively. Right panel: 95\% confidence Bayesian upper limits set on the angular power spectrum $(C_\ell^{95\%})^{1/2}$ with units sr$^{-1}$ for all frequencies and all harmonic modes are shown. The horizontal gaps are the notched frequencies. }
    \label{fig:upper_limit}
\end{figure*}

Though the primordial background is expected to be isotropic, there are possibilities of anisotropies being produced later in the Universe. In the next step, we analyze the data to find the signatures of narrow band anisotropic SGWB. We invert the Fisher information matrix and obtain the clean unbiased estimators for the angular power spectrum using Eqs.~(\ref{ang_power_spectra}), (\ref{eq:C_l_Cov}), (\ref{eq:ang_power_spectra_snr}). 

We define the angular power spectrum in terms of the dirty map as
\begin{equation}\label{eq:dirty_Cl}
    \hat{D}_\ell (f) = \frac{1}{2\ell+1}\sum_m |X_{\ell m}(f)|^2\,,
\end{equation}
whose expected value can be written as a convolution equation as
\begin{equation}
    \langle\hat{D}_\ell\rangle_{N,U}\approx \sum_{l'}A_{\ell\ell'}\left[\left(\frac{2\pi^2f^3}{3H^2_0}\right)^{-2}\,C_{\ell'}+4\pi\bar{\mathcal{P}}^2\delta_{\ell'0}\right]+N_\ell\,,
\end{equation}
where $A_{\ell\ell'}=\frac{1}{2\ell+1}\sum_{mm'}|\Gamma_{lm,l'm'}|^2$ can be called a bias or response matrix in analogy to CMB analysis~\cite{PhysRevD.70.103002} that represents the correlation between modes, and $N_\ell=\frac{1}{2\ell+1}\sum_m \Gamma_{\ell m,\ell m}$ is the noise angular power spectrum (see Appendix~\ref{sec:dirty_Cl} for details).
We use $\mathbf{A}$ matrix to qualitatively measure the detector response to two harmonic modes $\ell$ and $\ell'$ at a frequency $f$ and the correlation between them. In Fig.~\ref{fig:covMatDemo}, the examples for the frequencies 40.5 and 200.5 Hz are shown.  We use a bias or response matrix to find the suitable angular scale given a frequency with a maximum up to $\ell_{\text{max}}=15$ to estimate the significance of detection and the upper limits. The baseline's sensitivity to higher modes increases as frequency increases, and beyond the diffraction limit, it again starts to decrease but has significant sensitivity for around 3-4 extra modes~\cite{Neil_2001}. Hence, $\ell_{\text{max}}$ is defined as the point where the response fall by $10^{-5}$ of the maximum. This value is chosen such that the modes lower than the diffraction limit have a sensitivity of the same order. The upper limit of $\ell_{\text{max}}=15$ is chosen based on the most sensitive frequency of the O3 HL baseline, around $\sim$200 Hz~\cite{asaf_lvk}. Also, the harmonic modes are severely correlated as shown in Fig.~\ref{fig:covMatDemo} through correlation matrix and singular value spectrum. The inversion of the Fisher matrix $\mathbf{\Gamma}$ is performed by setting the condition number to 100 for ignoring insensitive modes\footnote{The condition number is chosen to be 100 such that we do not discard too many modes, on the other hand the noise fluctuations should not increase much. Also, as shown in the singular value plot (right panel of Fig.~\ref{fig:covMatDemo}), the first knee structure is covered by condition number of 100. If we increase it to 1000, many more insensitive modes will start contributing, which we want to avoid. If we lower it to 10, too few modes will contribute, which is also not desirable. In fact, we ran the analysis for condition number 10 and 1000 as well and did not find any outliers. In practice, the choice may depend on the kind of source distribution one is looking for and may need to run the analysis for different values of the condition number.}. 

\begin{figure*}
    \centering
    \includegraphics[width=\textwidth
    ]{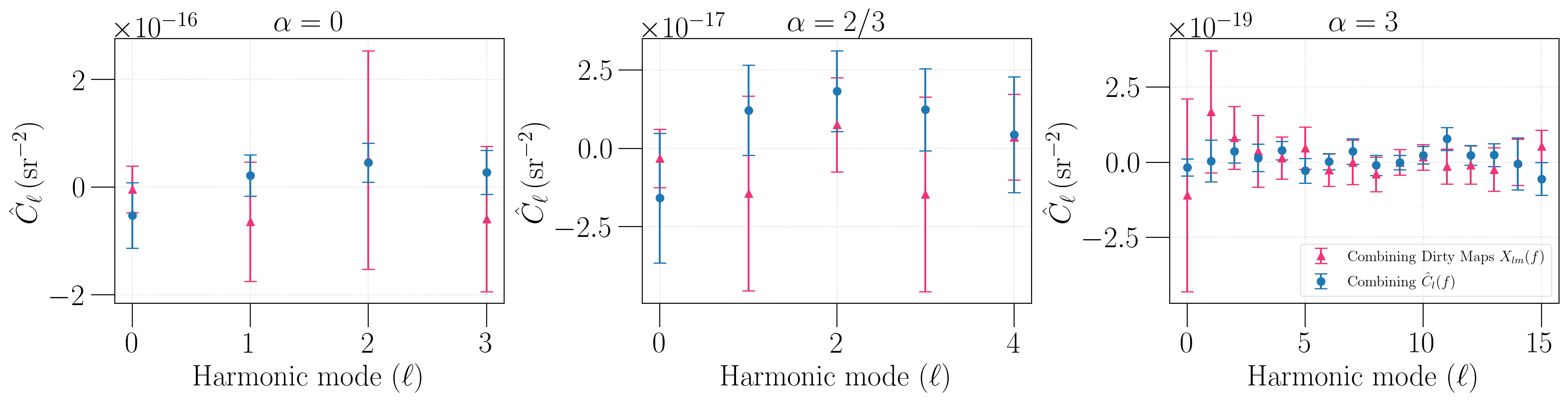}
    \caption{The ML estimators and their uncertainties (two sigma) for the broadband angular power spectrum with $f_{\text{ref}}=25$ Hz and spectral index $\alpha={0,2/3,3}$ obtained from combining the narrow band angular power spectrum (orange) and compared with the standard method (blue) of combing the dirty maps for the SpH coefficients.}
    \label{fig:nbr2bbr}
\end{figure*}

In the next step, we find the detection significance of each mode against the null hypothesis (that is, no signal presence, noise only). The null distribution is obtained using random time shift technique~\cite{asaf_lvk}: the time-series data from one detector is shifted by $\sim 1-2$ s with respect to the other detector's data to remove any coherence, and the estimators of the angular power spectra are computed by repeating the identical procedure. The distribution of SNRs is shown in Fig.~\ref{fig:histogram}, where the frequency samples are treated as independent samples for each mode. We note that the null distribution of SNR, the estimators of the angular power spectra, is no longer Gaussian. We find the best-fit DOF for the approximated PDF by minimizing the inverse-noise weighted mean-squared-error fitting to the time-shifted data (see Appendix~\ref{sec:mse}). Then, we calculate the SNR threshold for a global p value of 5\%. The approximated PDF is consistent with the time-shifted data within three-sigma Poisson error bars. The zero-lag data is also broadly consistent with the time-shifted data. A few candidates in zero-lag data have SNR above the threshold, which can be due to statistical fluctuations or coupling between the harmonic modes. This interpretation seems reasonable because when we plot the joint histogram of all samples, the zero-lag data is observed to be consistent with the time-shifted data (see Fig.~\ref{fig:upper_limit}). While the p value obtained from Fig.~\ref{fig:upper_limit} will not be precise, it will give an idea when to consider an observed multipole moment as a potential outlier. A more rigorous p value can then be obtained by going back to Fig.~\ref{fig:histogram} for the specific multipole. We set the Bayesian upper limit with $95\%$ confidence marginalized over calibration uncertainty. The upper limits set on $C_\ell^{1/2}$ for each $\ell$ as a function of frequency are presented in Fig.~\ref{fig:upper_limit}. The analytical expression of the likelihood for $C_\ell$ is nontrivial due to colored correlated noise. Hence the Bayesian posteriors are constructed using Monte Carlo sampling from multivariate normal distribution for SpH coefficients and marginalized over calibration uncertainty. The upper limit lies in range $3.0\times10^{-9}-0.73$ sr$^{-1}$. 

We conclude this part after comparing our results with predictions from a theoretical model for anisotropy of the SGWB created by the population of compact binary mergers. \citet{Capurri_2021} predicts the isotropic term and angular power spectrum for anisotropy at the reference frequency 65 Hz after removing the foreground sources detected by aLIGO/Virgo detectors. For the harmonic modes $2\leq\ell\leq4$, the isotropic term and the angular power spectrum lie in the range $\bar{\Omega}_{\textrm{GW}}\sim 10^{-9}$  and $1.5\times10^{-7}\lesssim\frac{\ell(\ell+1)\tilde{C}_\ell}{2\pi}\lesssim10^{-6}$ for the binary black hole population. Our upper limits are related to the predicted quantities as $C_\ell^{1/2}=\frac{\bar{\Omega}_{\textrm{GW}}}{4\pi}\sqrt{\tilde{C}_\ell}$.
As expected, due to the dominated detector noise, our upper limits at 65 Hz, $8.4\times10^{-8}\leq C_\ell^{1/2}\leq7.6\times10^{-8}$ sr$^{-1}$ are higher than the predicted anisotropy $3.2\times10^{-14}\lesssim C_\ell^{1/2}\lesssim 4.5\times10^{-14}$ sr$^{-1}$ by several of orders. As the detector network grows with improved sensitivity, we may be able to detect~\cite{Mentasti_2021} the anisotropy with the third-generation detectors such as the Einstein Telescope~\cite{ET}.

\subsection{Broadband higher multipoles}

 The angular power spectrum for a broadband SGWB with known spectral index can be obtained by adding narrow band dirty maps for the SpH coefficients and its covariance matrices with suitable weights, as was done in~\citet{O3directional}[see Eqs. (\ref{eq:Dirty_map_BBR}), (\ref{eq_fisher_sph_BBR})]. We will refer this method as ``standard method" further. It may also be of interest to derive these estimators for broadband angular power spectrum $\mathbf{C}$ using the narrow band angular power spectrum's estimators.

We attempt this problem with the assumption of additive multivariate Gaussian noise in the harmonic domain, i.e., $\mathbf{\hat{C}}(f)\equiv\hat{C}_\ell(f)$ is a random vector which obeys the multivariate Gaussian distribution\footnote{The Gaussian noise assumption may be valid at higher harmonic modes due to central limit theorem. The exact analytical expression for the PDF can be further explored. Here, we prescribe the treatment of combining narrow band estimators to obtain the estimators for a broadband background.} with covariance matrix $\mathbf{\Sigma}(f)$, the joint PDF (log-likelihood) using the estimator $\mathbf{\hat{C}}_\ell(f)$ from multiple frequency bins is given as

\begin{widetext}
\begin{eqnarray}
    -2\,\ln{L}= \sum_f \left[\mathbf{\hat{C}}(f)-w^2(f)\,\mathbf{C}\right]\cdot\mathbf{\Sigma}^{-1}(f)   \cdot \left[\mathbf{\hat{C}}(f)-w^2(f)\,\mathbf{C}\right]\,,
\end{eqnarray}
where $w(f) = (f/f_{\text{ref}})^{\alpha-3}$ is the usual weight for each frequency bin. The ML estimator for the broadband angular power spectrum $\mathbf{C}$ in terms of narrow band estimators is given as
\begin{equation}
    \hat{\mathbf{C}}=\left(\sum_f w^4(f)\,\mathbf{I}\cdot\mathbf{\Sigma}^{-1}(f)\cdot\mathbf{I}\right)^{-1}\left(\sum_f w^2(f)\,\mathbf{I}\cdot\mathbf{\Sigma}^{-1}(f)\cdot\mathbf{\mathbf{\hat{C}}}(f)\right)\,,
\end{equation}
and its covariance matrix as
\begin{equation}
\left(\sum_f w^4(f)\,\mathbf{I}\cdot\mathbf{\Sigma}^{-1}(f)\cdot\mathbf{I}\right)^{-1}\,,
\end{equation}
where $\mathbf{I}$ is a square identity matrix with dimension $\ell_{\text{max}}$.
\end{widetext}

In Fig.~\ref{fig:nbr2bbr}, the broadband estimators for three spectral indices $\alpha=[0,2/3,3]$ are presented for the HL baseline only. We note that the estimated broadband angular power spectra obtained by combining narrow band angular power spectra lie within $\sim 2$-sigma error bars compared to the estimators obtained by combining the narrow band dirty maps. The error bars obtained by our method are observed to be smaller at lower harmonics (say $\ell=[0,5]$), but for harmonics $\ell=[6,15]$ and spectral index $\alpha=3$, the error bars are quite in agreement. The difference in the estimators and their error bars are expected to be the cumulative effect of the Gaussian approximation and the nontrivial modification of singular modes during the regularization. We have presented a scheme to combine the estimators. However, the multivariate Gaussian likelihood may not be a good approximation at lower $\ell$ values. It could be a good approximation at higher $\ell$ due to increased DOF because of the central limit theorem. This issue is left for further exploration.



\section{Conclusions}\label{sec:conclusion}

We present the first narrow band (1 Hz) angular power spectra $C_\ell(f)$ by analyzing data from the third observing run of the Advanced LIGO and Advanced Virgo detectors. It aims to probe extended narrow band sources of the SGWB. We do not find any significant signature of narrow band SGWB. We set the $95\%$ confidence Bayesian upper limits on the narrow band $C_\ell(f)$, which may be helpful to constrain astrophysical and cosmological models.

We start by deriving the estimator for the isotropic component ($\ell_{\text{max}}=0$) for a narrow band and broadband statistically isotropic background and find the results to be consistent with the standard isotropic search. Then we estimate the narrow band angular power spectra, which involve the inversion of the Fisher-information matrix (the deconvolution problem). Here, we have implemented the SVD-based regularization scheme with a reasonable choice of the condition number. We also note that the detector's sensitivity to higher modes increases as the signal frequency increases. Hence, we set a threshold on analyzed harmonic modes for a given frequency (diffraction limit) before performing an inversion.

The clean angular power spectrum estimators are still correlated, which creates hurdles in obtaining an analytical expression for the likelihood function. Since we used the SNR of $C_\ell(f)$ as the detection statistic, we attempted to get the PDF for SNR with the assumption that no signal is present and the modes are uncorrelated. The approximated distribution is found to be consistent with the noise realization obtained by the timeshift method (but with modified DOF obtained by fitting the realization). There is a possibility of improvement in approximating the PDF, which is left for future exploration. In the absence of any significant signal, we set Bayesian upper limits on the angular power spectra with $95\%$ confidence lying in the range $C_\ell^{1/2}\leq(3.1\times10^{-9}-0.76) \text{ sr}^{-1}$. 

We have also presented a method to obtain the angular power spectrum for a broadband stochastic background by combining narrow band estimators. The optimal combination requires an analytical expression for the likelihood. We have assumed it to be the multivariate Gaussian distribution. This approximation may be accurate for higher harmonic modes due to the central limit theorem. The broadband angular power spectra obtained by combining narrow band angular power spectra lie within 2-sigma error bars compared to the estimators obtained by combining the narrow band dirty maps. 

The monopole estimator and two-point correlation function provide a complete description of a statistically isotropic Gaussian background which may be the case for the early Universe background and the background created by distant extragalactic sources. The SGWB from galactic sources may violate this isotropic condition, but performing a targeted search for the galactic plane~\cite{targeted_galPlane} and subtracting it from the observed sky may provide a practical solution. It was predicted that the angular power spectrum for the astrophysical background is non-Gaussian, particularly for late-Universe
sources, due to nonlinear dynamics of gravitational clustering~\cite{jenkins_thesis,BERNARDEAU20021}. In this case, analyzing higher-order statistics, such as bispectrum, trispectrum, etc., can provide more complete description. We leave this work for future exploration.

During the estimation of the uncertainty in the angular power spectrum, we have ignored the contribution from the shot noise~\cite{PhysRevD.100.063508,jenkins_thesis}. The shot noise arises from the transient nature of the source and may not be important for the continuous GW sources, e.g., GWs due mountains on the pulsars. Since the current GW detectors are operating in detector noise dominating regime, the treatment of the shot noise is left safely for future work. 

The techniques and results presented here can not only help constrain the theoretical models that predict the angular power spectrum for the source, but they can also be useful for cross-correlation studies involving, e.g., sky maps from electromagnetic observations such as cosmic microwave background anisotropy, galaxy count surveys, and gravitational lensing surveys.



\begin{acknowledgments}

The authors thank Joseph D. Romano for carefully reading the manuscript and providing valuable comments. This work significantly benefitted from the interactions with the Stochastic Working Group of the LIGO-Virgo-KAGRA Scientific Collaboration. J. S is supported by a Actions de Recherche Concertées (ARC) grant. S. M. acknowledges support from the Department of Science and Technology (DST), Ministry of Science and Technology, India, provided under the Swarna Jayanti Fellowships scheme. 

This material is based upon work supported by NSF's LIGO Laboratory, which is a major facility fully funded by the National Science Foundation. The authors are grateful for computational resources provided by the LIGO Laboratory (CIT) supported by National Science Foundation Grants No. PHY-0757058 and No. PHY-0823459, and Inter-University Center for Astronomy and Astrophysics (Sarathi). This research has made use of data or software obtained from the Gravitational Wave Open Science Center~\cite{gwosc}, a service of LIGO Laboratory, the LIGO Scientific Collaboration, the Virgo Collaboration, and KAGRA. LIGO Laboratory and Advanced LIGO are funded by the United States National Science Foundation (NSF) as well as the Science and Technology Facilities Council (STFC) of the United Kingdom, the Max-Planck-Society (MPS), and the State of Niedersachsen/Germany for support of the construction of Advanced LIGO and construction and operation of the GEO600 detector. Additional support for Advanced LIGO was provided by the Australian Research Council. Virgo is funded, through the European Gravitational Observatory (EGO), by the French Centre National de Recherche Scientifique (CNRS), the Italian Istituto Nazionale di Fisica Nucleare (INFN) and the Dutch Nikhef, with contributions by institutions from Belgium, Germany, Greece, Hungary, Ireland, Japan, Monaco, Poland, Portugal, Spain. The construction and operation of KAGRA are funded by Ministry of Education, Culture, Sports, Science and Technology (MEXT), and Japan Society for the Promotion of Science (JSPS), National Research Foundation (NRF) and Ministry of Science and ICT (MSIT) in Korea, Academia Sinica (AS) and the Ministry of Science and Technology (MoST) in Taiwan. This article has a LIGO Document No. LIGO-P2300030.

We used numerous software packages such as {\tt numpy}~\cite{5725236}, {\tt scipy}~\cite{Virtanen2020}, {\tt PyStoch}~\cite{pystoch,pystoch_sph} and
{\tt matplotlib}~\cite{Hunter:2007} in this work. 

\end{acknowledgments}



\appendix
\begin{widetext}
\section{Useful Derivations}
The dirty map in SpH basis can be written as~\cite{eric_sph}
\begin{equation}\label{eq:dirty_map}
   \mathbf{X}=\mathbf{\Gamma}\cdot\bm{\mathcal{P}}+\mathbf{n}\,,\quad\quad X_{\ell m} = \Gamma_{\ell m,\ell' m'}\mathcal{P}_{\ell'm'}+n_{\ell m}\,,
\end{equation}
where $\mathbf{n}$ and $\bm{\mathcal{P}}$ are column matrices having the SpH coefficients of the additive Gaussian noise $\{n_{\ell m}\}$, and SGWB sky map $\{\mathcal{P}_{\ell m}\}$, respectively as elements.
\begin{eqnarray}
    \langle\mathbf{n}\rangle_N=0\quad &\mathrm{and}& \quad\langle\mathbf{n}\mathbf{n}^\dagger\rangle_N\approx\mathbf{\Gamma}\,,\\
    \langle\mathbf{X}\rangle_N=\mathbf{\Gamma}\cdot\bm{\mathcal{P}}\quad &\mathrm{and}& \quad\langle\mathbf{X}\mathbf{X}^\dagger\rangle_N-\langle\mathbf{X}\rangle_N\langle\mathbf{X}^\dagger\rangle_N\approx\mathbf{\Gamma}\,,
\end{eqnarray}
where $\langle\cdot\rangle_N$ is the ensemble average over noise realizations. The additive noise $\mathbf{n}$ is expected to follow a multivariate Gaussian distribution. The statistical properties of the clean map are given as
\begin{equation}\label{eq:clean_map}
    \langle\bm{\hat{\mathcal{P}}}\rangle_N=(\mathbf{\Gamma}_R^{-1}\mathbf{\Gamma})\cdot\bm{\mathcal{P}}\quad \mathrm{and}\quad\langle\bm{\hat{\mathcal{P}}}\bm{\hat{\mathcal{P}}}^\dagger\rangle_N-\langle\bm{\hat{\mathcal{P}}}\rangle_N\langle\bm{\hat{\mathcal{P}}}^\dagger\rangle_N\approx\mathbf{\Gamma}_R^{-1}\mathbf{\Gamma}\mathbf{\Gamma}_R^{-1}\,.
\end{equation}
Note that the regularized clean map is a ``biased" estimator of the true SGWB SpH coefficient, and harmonic modes are correlated. Then, the unbiased angular power spectrum (see Eq.~(\ref{ang_power_spectra})) can be obtained similarly as in~\citet{eric_sph} by subtracting the bias due to noise covariance. 

\subsection{Covariance matrix for $\hat{C}_\ell$}\label{sec:covMatDerivation}
\begin{equation}
\Sigma_{\ell\ell'} = \langle\hat{C}_\ell\,\hat{C}_{\ell'}\rangle_{N,U}-\langle\hat{C}_\ell\rangle_{N,U}\langle\hat{C}_{\ell'}\rangle_{N,U}\,.
\end{equation}
We simplify the first term in the above expression (define $\mathcal{N}=\left(\frac{2\pi^2f^3}{3H^2_0})\right)^{2}$):
\begin{eqnarray}
  \langle\hat{C}_\ell\,\hat{C}_{\ell'}\rangle_{N,U}=\frac{\mathcal{N}^2}{(2\ell+1)(2\ell'+1)}&\langle&\sum_{m=-\ell}^{\ell}\bigg[|\hat{\mathcal{P}}_{\ell m}|^2-(\mathbf{\Gamma}_R^{-1}\cdot\mathbf{\Gamma}\cdot\mathbf{\Gamma}_R^{-1})_{\ell m,\ell m}\bigg]\sum_{m'=-\ell'}^{\ell'}\bigg[|\hat{\mathcal{P}}_{\ell' m'}|^2-(\mathbf{\Gamma}_R^{-1}\cdot\mathbf{\Gamma}\cdot\mathbf{\Gamma}_R^{-1})_{\ell' m',\ell' m'}\bigg]\rangle_{N,U}\,,\nonumber\\
 \frac{(2\ell+1)(2\ell'+1)}{\mathcal{N}^2}\langle\hat{C}_\ell\,\hat{C}_{\ell'}\rangle &=&\langle\sum_{m=-\ell}^{\ell}\sum_{m'=-\ell'}^{\ell'}|\hat{\mathcal{P}}_{\ell m}|^2|\hat{\mathcal{P}}_{\ell' m'}|^2\rangle_{N,U} - \sum_{m=-\ell}^{\ell}(\mathbf{\Gamma}_R^{-1}\cdot\mathbf{\Gamma}\cdot\mathbf{\Gamma}_R^{-1})_{\ell m,\ell m}\sum_{m'=-\ell'}^{\ell'}\langle|\hat{\mathcal{P}}_{\ell' m'}|^2\rangle_{N,U}\nonumber\\
 && - \sum_{m'=-\ell'}^{\ell'}(\mathbf{\Gamma}_R^{-1}\cdot\mathbf{\Gamma}\cdot\mathbf{\Gamma}_R^{-1})_{\ell ' m',\ell' m'}\sum_{m=-\ell}^{\ell}\langle|\hat{\mathcal{P}}_{\ell m}|^2\rangle_{N,U}\nonumber\\
 && +\sum_{m=-\ell}^{\ell}(\mathbf{\Gamma}_R^{-1}\cdot\mathbf{\Gamma}\cdot\mathbf{\Gamma}_R^{-1})_{\ell m,\ell m} \sum_{m'=-\ell'}^{\ell'}(\mathbf{\Gamma}_R^{-1}\cdot\mathbf{\Gamma}\cdot\mathbf{\Gamma}_R^{-1})_{\ell ' m',\ell' m'}\,.
\end{eqnarray}
Simplifying the second term,
\begin{eqnarray}
\frac{(2\ell+1)(2\ell'+1)}{\mathcal{N}^2}\langle\hat{C}_\ell\rangle_{N,U}\langle\hat{C}_{\ell'}\rangle_{N,U} &=&\sum_{m=-\ell}^{\ell}\langle|\hat{\mathcal{P}}_{\ell m}|^2\rangle_{N,U}\sum_{m'=-\ell'}^{\ell'}\langle|\hat{\mathcal{P}}_{\ell' m'}|^2\rangle_{N,U} - \sum_{m'=-\ell'}^{\ell'}\langle|\hat{\mathcal{P}}_{\ell' m'}|^2\rangle_{N,U}\sum_{m=-\ell}^{\ell}(\mathbf{\Gamma}_R^{-1}\cdot\mathbf{\Gamma}\cdot\mathbf{\Gamma}_R^{-1})_{\ell m,\ell m}\nonumber\\
&&- \sum_{m=-\ell}^{\ell}\langle|\hat{\mathcal{P}}_{\ell m}|^2\rangle_{N,U}\sum_{m'=-\ell'}^{\ell'}(\mathbf{\Gamma}_R^{-1}\cdot\mathbf{\Gamma}\cdot\mathbf{\Gamma}_R^{-1})_{\ell' m',\ell' m'}\nonumber\\
&& +\sum_{m=-\ell}^{\ell}(\mathbf{\Gamma}_R^{-1}\cdot\mathbf{\Gamma}\cdot\mathbf{\Gamma}_R^{-1})_{\ell m,\ell m} \sum_{m'=-\ell'}^{\ell'}(\mathbf{\Gamma}_R^{-1}\cdot\mathbf{\Gamma}\cdot\mathbf{\Gamma}_R^{-1})_{\ell ' m',\ell' m'}\,.
\end{eqnarray}
Hence,
\begin{equation}\label{eq:cov1}
\frac{(2\ell+1)(2\ell'+1)}{\mathcal{N}^2}\,\Sigma_{\ell\ell'} =\langle\sum_{m=-\ell}^{\ell}\sum_{m'=-\ell'}^{\ell'}|\hat{\mathcal{P}}_{\ell m}|^2|\hat{\mathcal{P}}_{\ell' m'}|^2\rangle_{N,U}-\sum_{m=-\ell}^{\ell}\langle|\hat{\mathcal{P}}_{\ell m}|^2\rangle_{N,U}\sum_{m'=-\ell'}^{\ell'}\langle|\hat{\mathcal{P}}_{\ell' m'}|^2\rangle_{N,U} \,. 
\end{equation}
Then simplifying the first term in the rhs of the above equation (in the weak signal limit),
\begin{eqnarray}\label{eq:cov2}
\langle\sum_{m=-\ell}^{\ell}\sum_{m'=-\ell'}^{\ell'}|\hat{\mathcal{P}}_{\ell m}|^2|\hat{\mathcal{P}}_{\ell' m'}|^2\rangle_{N,U}&=&\langle\sum_{m=-\ell}^{\ell}\sum_{m'=-\ell'}^{\ell'}|(\hat{\mathcal{P}}_{\ell m}-\mathcal{P}_{\ell m})+\mathcal{P}_{\ell m}|^2|\hat{\mathcal{P}}_{\ell' m'}-\mathcal{P}_{\ell'm'}+\mathcal{P}_{\ell' m'}|^2\rangle_{N,U}\\
    &\approx&\sum_{mm'}\langle|\hat{\mathcal{P}}_{\ell m}-\mathcal{P}_{\ell m}|^2|\hat{\mathcal{P}}_{\ell' m'}-\mathcal{P}_{\ell' m'}|^2\rangle_{N,U}+\sum_{mm'}\langle|\mathcal{P}_{\ell m}|^2\mathcal{P}_{\ell' m'}|^2\rangle_U\nonumber\\
    &+&\sum_{mm'}(\mathbf{\Gamma}_R^{-1}\cdot\mathbf{\Gamma}\cdot\mathbf{\Gamma}_R^{-1})_{\ell ' m',\ell' m'}\langle|\mathcal{P}_{\ell m}|^2\rangle_U+\sum_{mm'}(\mathbf{\Gamma}_R^{-1}\cdot\mathbf{\Gamma}\cdot\mathbf{\Gamma}_R^{-1})_{\ell m,\ell m}\langle|\mathcal{P}_{\ell' m'}|^2\rangle_U\nonumber\\
    &+&\sum_{mm'}4\, \mathcal{R}\left[(\mathbf{\Gamma}_R^{-1}\cdot\mathbf{\Gamma}\cdot\mathbf{\Gamma}_R^{-1})_{\ell m,\ell' m'}\langle \mathcal{P}_{\ell m}\mathcal{P}^{*}_{\ell' m'}\rangle_U\right]\,.
\end{eqnarray}
Also, using Wick's theorem for the Gaussian random variables~\cite{Zee:2003mt},
\begin{equation}
    \sum_{mm'}\langle|\hat{\mathcal{P}}_{\ell m}-\mathcal{P}_{\ell m}|^2|\hat{\mathcal{P}}_{\ell' m'}-\mathcal{P}_{\ell' m'}|^2\rangle_{N,U}\approx\sum_{mm'}(\mathbf{\Gamma}_R^{-1}\cdot\mathbf{\Gamma}\cdot\mathbf{\Gamma}_R^{-1})_{\ell m,\ell m}(\mathbf{\Gamma}_R^{-1}\cdot\mathbf{\Gamma}\cdot\mathbf{\Gamma}_R^{-1})_{\ell' m',\ell' m'}+\sum_{mm'}2|(\mathbf{\Gamma}_R^{-1}\cdot\mathbf{\Gamma}\cdot\mathbf{\Gamma}_R^{-1})_{\ell m,\ell' m'}|^2\,.
\end{equation}

Solving for the second part of Eq.~(\ref{eq:cov1}) and using Eq. (\ref{eq:clean_map}),

\begin{eqnarray}
    \sum_{m=-\ell}^{\ell}\langle|\hat{\mathcal{P}}_{\ell m}|^2\rangle_{N,U}\sum_{m'=-\ell'}^{\ell'}\langle|\hat{\mathcal{P}}_{\ell' m'}|^2\rangle_{N,U} =&&\sum_{m=-\ell}^{\ell}\left[\langle|\mathcal{P}_{\ell m}|^2\rangle_{U}+(\mathbf{\Gamma}_R^{-1}\cdot\mathbf{\Gamma}\cdot\mathbf{\Gamma}_R^{-1})_{\ell m,\ell m}\right]\nonumber\\
    &&\sum_{m'=-\ell'
    }^{\ell'}\left[\langle|\mathcal{P}_{\ell' m'}|^2\rangle_{U}+(\mathbf{\Gamma}_R^{-1}\cdot\mathbf{\Gamma}\cdot\mathbf{\Gamma}_R^{-1})_{\ell m,\ell' m'}\right]\,.
\end{eqnarray}

Putting all pieces together
\begin{eqnarray}
\frac{(2\ell+1)(2\ell'+1)}{\mathcal{N}^2}\,\Sigma_{\ell\ell'} &\approx&\sum_{mm'}\langle|\mathcal{P}_{\ell m}|^2\mathcal{P}_{\ell' m'}|^2\rangle_U-\sum_{m
    }\langle|\mathcal{P}_{\ell m}|^2\rangle_{U}\sum_{m
    }\langle|\mathcal{P}_{\ell' m'}|^2\rangle_{U}\nonumber\\
&+&\sum_{mm'}4\, \mathcal{R}\left[(\mathbf{\Gamma}_R^{-1}\cdot\mathbf{\Gamma}\cdot\mathbf{\Gamma}_R^{-1})_{\ell m,\ell' m'}\langle \mathcal{P}_{\ell m}\mathcal{P}^{*}_{\ell' m'}\rangle_U\right]+\sum_{mm'}2|(\mathbf{\Gamma}_R^{-1}\cdot\mathbf{\Gamma}\cdot\mathbf{\Gamma}_R^{-1})_{\ell m,\ell' m'}|^2\\
&\approx&(2\ell+1)^2\,\textrm{Var}[\mathcal{N}^{-1}\,\tilde{C}_\ell]_U\,\delta_{\ell\ell'}+\sum_m\delta_{\ell\ell'}4\, \mathcal{R}\left[(\mathbf{\Gamma}_R^{-1}\cdot\mathbf{\Gamma}\cdot\mathbf{\Gamma}_R^{-1})_{\ell m,\ell m}\right]\left[\mathcal{N}^{-1}\,C_\ell+4\pi\bar{\mathcal{P}}^2\delta_{\ell0}\right]\nonumber\\
&+&\sum_{mm'}2|(\mathbf{\Gamma}_R^{-1}\cdot\mathbf{\Gamma}\cdot\mathbf{\Gamma}_R^{-1})_{\ell m,\ell' m'}|^2\,.
\end{eqnarray}

If we assume that the signal is weak and the true angular power spectrum is negligible in comparison to the noise spectrum, then covariance
\begin{equation}
\frac{(2\ell+1)(2\ell'+1)}{\mathcal{N}^2}\,\,\Sigma_{\ell\ell'} \approx\sum_{mm'}2|(\mathbf{\Gamma}_R^{-1}\cdot\mathbf{\Gamma}\cdot\mathbf{\Gamma}_R^{-1})_{\ell m,\ell' m'}|^2\,.
\end{equation}

\subsection{Approximated PDF for SNR}\label{sec:apprxPDF}
Using Eqs.~(\ref{ang_power_spectra}) and (\ref{eq:snrpdfassum}),
\begin{equation}
    \hat{C}_\ell = \frac{\mathcal{N}}{2\ell+1}\sum_{m}\bigg[|\hat{\mathcal{P}}_{\ell m}|^2-\sigma_\ell^2\bigg]=\frac{\mathcal{N}\,\sigma_\ell^2}{2\ell+1}\sum_{m}\bigg[\frac{|\hat{\mathcal{P}}_{\ell m}|^2}{\sigma_\ell^2}-1\bigg]\,,
\end{equation}
\begin{equation}
    \hat{C}_\ell =\frac{\mathcal{N}\,\sigma_\ell^2}{2\ell+1}\sum_{m}\bigg[|\rho_{\ell m}|^2-1\bigg]\,,
\end{equation}
where $\rho_{\ell m}$ is the SNR of the SpH coefficient. Now, using Eq.~(\ref{eq:C_l_Cov}), the error bar can be written as
\begin{equation}
    \sigma_{C_\ell}=\sqrt{\frac{2\mathcal{N}^2}{(2\ell+1)^2}\sum_{m}\sigma_\ell^4}=\sqrt{\frac{2}{(2\ell+1)}}\,\mathcal{N}\,\sigma_\ell^2\,.
\end{equation}
Then the SNR is
\begin{equation}
\boxed{
    \rho_\ell = \sqrt{\frac{1}{2(2\ell+1)}}\sum_{m}\bigg[|\rho_{\ell m}|^2-1\bigg]\,.}
\end{equation}

The SNR $\rho_{\ell m}$ are normally distributed random variables in noise-only cases. Then, the PDF for the sum of the squares of the $k=2\ell+1$ normally distributed random variables is $\chi^2$ distributed with $k$ DOF:
\begin{equation}
    P\,\bigg(x=\sum_{m}|\rho_{\ell m}|^2\bigg)=\chi^2_{k}\bigg(x=\sum_{m}|\rho_{\ell m}|^2\bigg)\,.
\end{equation}
Then, the distribution of $\rho_\ell=\sqrt{\frac{1}{2k}}(x-k)$, from change of variable
\begin{eqnarray}
    P\,\bigg(y=\rho_\ell&=&\frac{(x-k)}{\sqrt{2k}})\bigg)\,dy = P\,(x(y)) \bigg|\frac{dx}
    {dy}\bigg| \,dy\nonumber\\
    &=&\sqrt{2k}\,\chi^2_{k}(x=\sqrt{2k}\,y+k)\,dy\,.
\end{eqnarray}

\subsection{Angular power spectrum using dirty map}\label{sec:dirty_Cl}

Let us define the angular power spectrum using the dirty map as
\begin{equation}
    \hat{D}_\ell = \frac{1}{2\ell+1}\sum_m |X_{\ell m}|^2\,.
\end{equation}
Then its expected value is given by
\begin{eqnarray}
    \langle\hat{D}_\ell\rangle_{N,U} &=& \frac{1}{2\ell+1}\sum_m \langle X^{*}_{\ell m}X_{\ell m}\rangle_{N,U}\\
    &=&\frac{1}{2\ell+1}\sum_m\sum_{l'm'}\sum_{l''m''}\Gamma^{*}_{lm,l'm'}\Gamma_{lm,l''m''}\langle \mathcal{P}_{l'm'}\mathcal{P}_{l''m''}\rangle_{N,U}+\frac{1}{2\ell+1}\sum_m\langle n^{*}_{\ell m}n_{\ell m}\rangle_{N,U}
    \\
     &=&\frac{1}{2\ell+1}\sum_m\sum_{l'm'}\sum_{l''m''}\Gamma^{*}_{lm,l'm'}\Gamma_{lm,l''m''}\langle \mathcal{P}_{l'm'}\mathcal{P}_{l''m''}\rangle_{U}+\frac{1}{2\ell+1}\sum_m\langle n^{*}_{\ell m}n_{\ell m}\rangle_{N}
    \\
   & \approx&\frac{1}{2\ell+1}\sum_m\sum_{l'm'}\sum_{l''m''}\Gamma^{*}_{lm,l'm'}\Gamma_{lm,l''m''}\,\left[\mathcal{N}^{-1}\,C_{\ell'}\delta_{\ell'\ell''}\delta_{m'm''}+4\pi\bar{\mathcal{P}}^2\delta_{\ell'\ell''}\delta_{\ell'0}\right]+\frac{1}{2\ell+1}\sum_m \Gamma_{\ell m,\ell m}\\ 
    &\approx& \frac{1}{2\ell+1}\sum_m\sum_{l'm'}\Gamma^{*}_{lm,l'm'}\Gamma^{*}_{lm,l'm'}\,\left[\mathcal{N}^{-1}C_{\ell'}+4\pi\bar{\mathcal{P}}^2\delta_{\ell'0}\right]+\frac{1}{2\ell+1}\sum_m \Gamma_{\ell m,\ell m}\\
   &\approx& \sum_{l'}A_{\ell\ell'}\left[\mathcal{N}^{-1}\,C_{\ell'}+4\pi\bar{\mathcal{P}}^2\delta_{\ell'0}\right]+N_\ell\,.
\end{eqnarray}
where $A_{\ell\ell'}=\frac{1}{2\ell+1}\sum_{mm'}|\Gamma_{lm,l'm'}|^2$ can be called a bias matrix in analogy to CMB analysis~\cite{PhysRevD.70.103002} that represents the correlation between modes, and $N_\ell=\frac{1}{2\ell+1}\sum_m \Gamma_{\ell m,\ell m}$ is the noise angular power spectrum.

\subsection{Weighted mean squared error}\label{sec:mse}

We fit the histogram from time-shifted data and find effective DOF by minimizing the weighted mean squared error (MSE) defined as
\begin{equation}
    \text{MSE} = \frac{\sum_{i=1}^{n_{\text{bins}}}\,w_i\,(N_i-N_{\text{total}}\,P(y_i))}{\sum_{i=1}^{n_{\text{bins}}}\,w_i}\,,,
\end{equation}
where $n_{\text{bins}}$ are the numbers of bins in the histogram, $N_i$ is the number of samples having statistic value between $y_i-dy/2$ and $y_i+dy/2$, $N_{\text{total}}$ is the total number of the samples to fit and $w_i=N_i^{-1}$ is the weight for each bin to give more weight to the tail of the histogram.
\end{widetext}


\bibliography{reference}

\begin{thebibliography}{52}%
\makeatletter
\providecommand \@ifxundefined [1]{%
 \@ifx{#1\undefined}
}%
\providecommand \@ifnum [1]{%
 \ifnum #1\expandafter \@firstoftwo
 \else \expandafter \@secondoftwo
 \fi
}%
\providecommand \@ifx [1]{%
 \ifx #1\expandafter \@firstoftwo
 \else \expandafter \@secondoftwo
 \fi
}%
\providecommand \natexlab [1]{#1}%
\providecommand \enquote  [1]{``#1''}%
\providecommand \bibnamefont  [1]{#1}%
\providecommand \bibfnamefont [1]{#1}%
\providecommand \citenamefont [1]{#1}%
\providecommand \href@noop [0]{\@secondoftwo}%
\providecommand \href [0]{\begingroup \@sanitize@url \@href}%
\providecommand \@href[1]{\@@startlink{#1}\@@href}%
\providecommand \@@href[1]{\endgroup#1\@@endlink}%
\providecommand \@sanitize@url [0]{\catcode `\\12\catcode `\$12\catcode
  `\&12\catcode `\#12\catcode `\^12\catcode `\_12\catcode `\%12\relax}%
\providecommand \@@startlink[1]{}%
\providecommand \@@endlink[0]{}%
\providecommand \url  [0]{\begingroup\@sanitize@url \@url }%
\providecommand \@url [1]{\endgroup\@href {#1}{\urlprefix }}%
\providecommand \urlprefix  [0]{URL }%
\providecommand \Eprint [0]{\href }%
\providecommand \doibase [0]{http://dx.doi.org/}%
\providecommand \selectlanguage [0]{\@gobble}%
\providecommand \bibinfo  [0]{\@secondoftwo}%
\providecommand \bibfield  [0]{\@secondoftwo}%
\providecommand \translation [1]{[#1]}%
\providecommand \BibitemOpen [0]{}%
\providecommand \bibitemStop [0]{}%
\providecommand \bibitemNoStop [0]{.\EOS\space}%
\providecommand \EOS [0]{\spacefactor3000\relax}%
\providecommand \BibitemShut  [1]{\csname bibitem#1\endcsname}%
\let\auto@bib@innerbib\@empty
\bibitem [{\citenamefont {Collaboration}\ \emph {et~al.}(2021)\citenamefont
  {Collaboration}, \citenamefont {the Virgo~Collaboration}, \citenamefont {the
  KAGRA~Collaboration} \emph {et~al.}}]{GWTC3}%
  \BibitemOpen
  \bibfield  {author} {\bibinfo {author} {\bibfnamefont {The LIGO~Scientific}\
  \bibnamefont {Collaboration}}, \bibinfo {author} {\bibnamefont {the
  Virgo~Collaboration}}, \bibinfo {author} {\bibnamefont {the
  KAGRA~Collaboration}},  \emph {et~al.},\ }\href@noop {} {\enquote {\bibinfo
  {title} {Gwtc-3: Compact binary coalescences observed by ligo and virgo
  during the second part of the third observing run},}\ } (\bibinfo {year}
  {2021}),\ \Eprint {http://arxiv.org/abs/2111.03606} {arXiv:2111.03606
  [gr-qc]} \BibitemShut {NoStop}%
\bibitem [{\citenamefont {Nitz}\ \emph {et~al.}(2021)\citenamefont {Nitz},
  \citenamefont {Capano}, \citenamefont {Kumar}, \citenamefont {Wang},
  \citenamefont {Kastha}, \citenamefont {Schäfer}, \citenamefont {Dhurkunde},\
  and\ \citenamefont {Cabero}}]{OGC4}%
  \BibitemOpen
  \bibfield  {author} {\bibinfo {author} {\bibfnamefont {Alexander~H.}\
  \bibnamefont {Nitz}}, \bibinfo {author} {\bibfnamefont {Collin~D.}\
  \bibnamefont {Capano}}, \bibinfo {author} {\bibfnamefont {Sumit}\
  \bibnamefont {Kumar}}, \bibinfo {author} {\bibfnamefont {Yi-Fan}\
  \bibnamefont {Wang}}, \bibinfo {author} {\bibfnamefont {Shilpa}\ \bibnamefont
  {Kastha}}, \bibinfo {author} {\bibfnamefont {Marlin}\ \bibnamefont
  {Schäfer}}, \bibinfo {author} {\bibfnamefont {Rahul}\ \bibnamefont
  {Dhurkunde}}, \ and\ \bibinfo {author} {\bibfnamefont {Miriam}\ \bibnamefont
  {Cabero}},\ }\href {\doibase 10.3847/1538-4357/ac1c03} {\enquote {\bibinfo
  {title} {3-ogc: Catalog of gravitational waves from compact-binary
  mergers},}\ } (\bibinfo {year} {2021})\BibitemShut {NoStop}%
\bibitem [{\citenamefont {Regimbau}(2011)}]{Regimbau_2011}%
  \BibitemOpen
  \bibfield  {author} {\bibinfo {author} {\bibfnamefont {Tania}\ \bibnamefont
  {Regimbau}},\ }\bibfield  {title} {\enquote {\bibinfo {title} {The
  astrophysical gravitational wave stochastic background},}\ }\href {\doibase
  10.1088/1674-4527/11/4/001} {\bibfield  {journal} {\bibinfo  {journal}
  {Research in Astronomy and Astrophysics}\ }\textbf {\bibinfo {volume} {11}},\
  \bibinfo {pages} {369--390} (\bibinfo {year} {2011})}\BibitemShut {NoStop}%
\bibitem [{\citenamefont {Regimbau}(2022)}]{Regimbau:2022mdu}%
  \BibitemOpen
  \bibfield  {author} {\bibinfo {author} {\bibfnamefont {Tania}\ \bibnamefont
  {Regimbau}},\ }\bibfield  {title} {\enquote {\bibinfo {title} {{The Quest for
  the Astrophysical Gravitational-Wave Background with Terrestrial
  Detectors}},}\ }\href {\doibase 10.3390/sym14020270} {\bibfield  {journal}
  {\bibinfo  {journal} {Symmetry}\ }\textbf {\bibinfo {volume} {14}},\ \bibinfo
  {pages} {270} (\bibinfo {year} {2022})}\BibitemShut {NoStop}%
\bibitem [{\citenamefont {Jenkins}\ and\ \citenamefont
  {Sakellariadou}(2018)}]{Jenkins:2018lvb}%
  \BibitemOpen
  \bibfield  {author} {\bibinfo {author} {\bibfnamefont {Alexander~C.}\
  \bibnamefont {Jenkins}}\ and\ \bibinfo {author} {\bibfnamefont {Mairi}\
  \bibnamefont {Sakellariadou}},\ }\bibfield  {title} {\enquote {\bibinfo
  {title} {{Anisotropies in the stochastic gravitational-wave background:
  Formalism and the cosmic string case}},}\ }\href {\doibase
  10.1103/PhysRevD.98.063509} {\bibfield  {journal} {\bibinfo  {journal} {Phys.
  Rev.}\ }\textbf {\bibinfo {volume} {D98}},\ \bibinfo {pages} {063509}
  (\bibinfo {year} {2018})},\ \Eprint {http://arxiv.org/abs/1802.06046}
  {arXiv:1802.06046 [astro-ph.CO]} \BibitemShut {NoStop}%
\bibitem [{\citenamefont {{Mazumder}}\ \emph {et~al.}(2014)\citenamefont
  {{Mazumder}}, \citenamefont {{Mitra}},\ and\ \citenamefont
  {{Dhurandhar}}}]{2014PhRvD..89h4076M}%
  \BibitemOpen
  \bibfield  {author} {\bibinfo {author} {\bibfnamefont {N.}~\bibnamefont
  {{Mazumder}}}, \bibinfo {author} {\bibfnamefont {S.}~\bibnamefont {{Mitra}}},
  \ and\ \bibinfo {author} {\bibfnamefont {S.}~\bibnamefont {{Dhurandhar}}},\
  }\bibfield  {title} {\enquote {\bibinfo {title} {{Astrophysical motivation
  for directed searches for a stochastic gravitational wave background}},}\
  }\href {\doibase 10.1103/PhysRevD.89.084076} {\bibfield  {journal} {\bibinfo
  {journal} {Phys. Rev. D}\ }\textbf {\bibinfo {volume} {89}},\ \bibinfo {eid}
  {084076} (\bibinfo {year} {2014})},\ \Eprint {http://arxiv.org/abs/1401.5898}
  {arXiv:1401.5898 [gr-qc]} \BibitemShut {NoStop}%
\bibitem [{\citenamefont {Cusin}\ \emph {et~al.}(2018)\citenamefont {Cusin},
  \citenamefont {Dvorkin}, \citenamefont {Pitrou},\ and\ \citenamefont
  {Uzan}}]{PhysRevLett.120.231101}%
  \BibitemOpen
  \bibfield  {author} {\bibinfo {author} {\bibfnamefont {Giulia}\ \bibnamefont
  {Cusin}}, \bibinfo {author} {\bibfnamefont {Irina}\ \bibnamefont {Dvorkin}},
  \bibinfo {author} {\bibfnamefont {Cyril}\ \bibnamefont {Pitrou}}, \ and\
  \bibinfo {author} {\bibfnamefont {Jean-Philippe}\ \bibnamefont {Uzan}},\
  }\bibfield  {title} {\enquote {\bibinfo {title} {First predictions of the
  angular power spectrum of the astrophysical gravitational wave background},}\
  }\href {\doibase 10.1103/PhysRevLett.120.231101} {\bibfield  {journal}
  {\bibinfo  {journal} {Phys. Rev. Lett.}\ }\textbf {\bibinfo {volume} {120}},\
  \bibinfo {pages} {231101} (\bibinfo {year} {2018})}\BibitemShut {NoStop}%
\bibitem [{\citenamefont {{Jenkins}}\ \emph {et~al.}(2018)\citenamefont
  {{Jenkins}}, \citenamefont {{Sakellariadou}}, \citenamefont {{Regimbau}},\
  and\ \citenamefont {{Slezak}}}]{jenkins_sakell_cbc_anisotropy}%
  \BibitemOpen
  \bibfield  {author} {\bibinfo {author} {\bibfnamefont {A.~C.}\ \bibnamefont
  {{Jenkins}}}, \bibinfo {author} {\bibfnamefont {M.}~\bibnamefont
  {{Sakellariadou}}}, \bibinfo {author} {\bibfnamefont {T.}~\bibnamefont
  {{Regimbau}}}, \ and\ \bibinfo {author} {\bibfnamefont {E.}~\bibnamefont
  {{Slezak}}},\ }\bibfield  {title} {\enquote {\bibinfo {title} {{Anisotropies
  in the astrophysical gravitational-wave background: Predictions for the
  detection of compact binaries by LIGO and Virgo}},}\ }\href {\doibase
  10.1103/PhysRevD.98.063501} {\bibfield  {journal} {\bibinfo  {journal}
  {\prd}\ }\textbf {\bibinfo {volume} {98}},\ \bibinfo {eid} {063501} (\bibinfo
  {year} {2018})},\ \Eprint {http://arxiv.org/abs/1806.01718}
  {arXiv:1806.01718} \BibitemShut {NoStop}%
\bibitem [{\citenamefont {{Rosado}}(2012)}]{2012PhRvD..86j4007R}%
  \BibitemOpen
  \bibfield  {author} {\bibinfo {author} {\bibfnamefont {Pablo~A.}\
  \bibnamefont {{Rosado}}},\ }\bibfield  {title} {\enquote {\bibinfo {title}
  {{Gravitational wave background from rotating neutron stars}},}\ }\href
  {\doibase 10.1103/PhysRevD.86.104007} {\bibfield  {journal} {\bibinfo
  {journal} {Phys. Rev. D}\ }\textbf {\bibinfo {volume} {86}},\ \bibinfo {eid}
  {104007} (\bibinfo {year} {2012})},\ \Eprint {http://arxiv.org/abs/1206.1330}
  {arXiv:1206.1330 [gr-qc]} \BibitemShut {NoStop}%
\bibitem [{\citenamefont {{Wu}}\ \emph {et~al.}(2013)\citenamefont {{Wu}},
  \citenamefont {{Mandic}},\ and\ \citenamefont
  {{Regimbau}}}]{2013PhRvD..87d2002W}%
  \BibitemOpen
  \bibfield  {author} {\bibinfo {author} {\bibfnamefont {Cheng-Jian}\
  \bibnamefont {{Wu}}}, \bibinfo {author} {\bibfnamefont {Vuk}\ \bibnamefont
  {{Mandic}}}, \ and\ \bibinfo {author} {\bibfnamefont {Tania}\ \bibnamefont
  {{Regimbau}}},\ }\bibfield  {title} {\enquote {\bibinfo {title}
  {{Accessibility of the stochastic gravitational wave background from
  magnetars to the interferometric gravitational wave detectors}},}\ }\href
  {\doibase 10.1103/PhysRevD.87.042002} {\bibfield  {journal} {\bibinfo
  {journal} {Phys. Rev. D}\ }\textbf {\bibinfo {volume} {87}},\ \bibinfo {eid}
  {042002} (\bibinfo {year} {2013})}\BibitemShut {NoStop}%
\bibitem [{\citenamefont {{Lasky}}\ \emph {et~al.}(2013)\citenamefont
  {{Lasky}}, \citenamefont {{Bennett}},\ and\ \citenamefont
  {{Melatos}}}]{2013PhRvD..87f3004L}%
  \BibitemOpen
  \bibfield  {author} {\bibinfo {author} {\bibfnamefont {Paul~D.}\ \bibnamefont
  {{Lasky}}}, \bibinfo {author} {\bibfnamefont {Mark~F.}\ \bibnamefont
  {{Bennett}}}, \ and\ \bibinfo {author} {\bibfnamefont {Andrew}\ \bibnamefont
  {{Melatos}}},\ }\bibfield  {title} {\enquote {\bibinfo {title} {{Stochastic
  gravitational wave background from hydrodynamic turbulence in differentially
  rotating neutron stars}},}\ }\href {\doibase 10.1103/PhysRevD.87.063004}
  {\bibfield  {journal} {\bibinfo  {journal} {Phys. Rev. D}\ }\textbf {\bibinfo
  {volume} {87}},\ \bibinfo {eid} {063004} (\bibinfo {year} {2013})},\ \Eprint
  {http://arxiv.org/abs/1302.6033} {arXiv:1302.6033 [astro-ph.HE]} \BibitemShut
  {NoStop}%
\bibitem [{\citenamefont {Cusin}\ \emph {et~al.}(2017)\citenamefont {Cusin},
  \citenamefont {Pitrou},\ and\ \citenamefont {Uzan}}]{PhysRevD.96.103019}%
  \BibitemOpen
  \bibfield  {author} {\bibinfo {author} {\bibfnamefont {Giulia}\ \bibnamefont
  {Cusin}}, \bibinfo {author} {\bibfnamefont {Cyril}\ \bibnamefont {Pitrou}}, \
  and\ \bibinfo {author} {\bibfnamefont {Jean-Philippe}\ \bibnamefont {Uzan}},\
  }\bibfield  {title} {\enquote {\bibinfo {title} {Anisotropy of the
  astrophysical gravitational wave background: Analytic expression of the
  angular power spectrum and correlation with cosmological observations},}\
  }\href {\doibase 10.1103/PhysRevD.96.103019} {\bibfield  {journal} {\bibinfo
  {journal} {Phys. Rev. D}\ }\textbf {\bibinfo {volume} {96}},\ \bibinfo
  {pages} {103019} (\bibinfo {year} {2017})}\BibitemShut {NoStop}%
\bibitem [{\citenamefont {Capurri}\ \emph {et~al.}(2021)\citenamefont
  {Capurri}, \citenamefont {Lapi}, \citenamefont {Baccigalupi}, \citenamefont
  {Boco}, \citenamefont {Scelfo},\ and\ \citenamefont
  {Ronconi}}]{Capurri_2021}%
  \BibitemOpen
  \bibfield  {author} {\bibinfo {author} {\bibfnamefont {Giulia}\ \bibnamefont
  {Capurri}}, \bibinfo {author} {\bibfnamefont {Andrea}\ \bibnamefont {Lapi}},
  \bibinfo {author} {\bibfnamefont {Carlo}\ \bibnamefont {Baccigalupi}},
  \bibinfo {author} {\bibfnamefont {Lumen}\ \bibnamefont {Boco}}, \bibinfo
  {author} {\bibfnamefont {Giulio}\ \bibnamefont {Scelfo}}, \ and\ \bibinfo
  {author} {\bibfnamefont {Tommaso}\ \bibnamefont {Ronconi}},\ }\bibfield
  {title} {\enquote {\bibinfo {title} {Intensity and anisotropies of the
  stochastic gravitational wave background from merging compact binaries in
  galaxies},}\ }\href {\doibase 10.1088/1475-7516/2021/11/032} {\bibfield
  {journal} {\bibinfo  {journal} {Journal of Cosmology and Astroparticle
  Physics}\ }\textbf {\bibinfo {volume} {2021}},\ \bibinfo {pages} {032}
  (\bibinfo {year} {2021})}\BibitemShut {NoStop}%
\bibitem [{\citenamefont {Bertacca}\ \emph {et~al.}(2020)\citenamefont
  {Bertacca}, \citenamefont {Ricciardone}, \citenamefont {Bellomo},
  \citenamefont {Jenkins}, \citenamefont {Matarrese}, \citenamefont
  {Raccanelli}, \citenamefont {Regimbau},\ and\ \citenamefont
  {Sakellariadou}}]{PhysRevD.101.103513}%
  \BibitemOpen
  \bibfield  {author} {\bibinfo {author} {\bibfnamefont {Daniele}\ \bibnamefont
  {Bertacca}}, \bibinfo {author} {\bibfnamefont {Angelo}\ \bibnamefont
  {Ricciardone}}, \bibinfo {author} {\bibfnamefont {Nicola}\ \bibnamefont
  {Bellomo}}, \bibinfo {author} {\bibfnamefont {Alexander~C.}\ \bibnamefont
  {Jenkins}}, \bibinfo {author} {\bibfnamefont {Sabino}\ \bibnamefont
  {Matarrese}}, \bibinfo {author} {\bibfnamefont {Alvise}\ \bibnamefont
  {Raccanelli}}, \bibinfo {author} {\bibfnamefont {Tania}\ \bibnamefont
  {Regimbau}}, \ and\ \bibinfo {author} {\bibfnamefont {Mairi}\ \bibnamefont
  {Sakellariadou}},\ }\bibfield  {title} {\enquote {\bibinfo {title}
  {Projection effects on the observed angular spectrum of the astrophysical
  stochastic gravitational wave background},}\ }\href {\doibase
  10.1103/PhysRevD.101.103513} {\bibfield  {journal} {\bibinfo  {journal}
  {Phys. Rev. D}\ }\textbf {\bibinfo {volume} {101}},\ \bibinfo {pages}
  {103513} (\bibinfo {year} {2020})}\BibitemShut {NoStop}%
\bibitem [{\citenamefont {Cusin}\ and\ \citenamefont
  {Tasinato}(2022)}]{Cusin_2022}%
  \BibitemOpen
  \bibfield  {author} {\bibinfo {author} {\bibfnamefont {Giulia}\ \bibnamefont
  {Cusin}}\ and\ \bibinfo {author} {\bibfnamefont {Gianmassimo}\ \bibnamefont
  {Tasinato}},\ }\bibfield  {title} {\enquote {\bibinfo {title} {Doppler
  boosting the stochastic gravitational wave background},}\ }\href {\doibase
  10.1088/1475-7516/2022/08/036} {\bibfield  {journal} {\bibinfo  {journal}
  {Journal of Cosmology and Astroparticle Physics}\ }\textbf {\bibinfo {volume}
  {2022}},\ \bibinfo {pages} {036} (\bibinfo {year} {2022})}\BibitemShut
  {NoStop}%
\bibitem [{\citenamefont {Chung}\ \emph {et~al.}(2022)\citenamefont {Chung},
  \citenamefont {Jenkins}, \citenamefont {Romano},\ and\ \citenamefont
  {Sakellariadou}}]{PhysRevD.106.082005}%
  \BibitemOpen
  \bibfield  {author} {\bibinfo {author} {\bibfnamefont {Adrian Ka-Wai}\
  \bibnamefont {Chung}}, \bibinfo {author} {\bibfnamefont {Alexander~C.}\
  \bibnamefont {Jenkins}}, \bibinfo {author} {\bibfnamefont {Joseph~D.}\
  \bibnamefont {Romano}}, \ and\ \bibinfo {author} {\bibfnamefont {Mairi}\
  \bibnamefont {Sakellariadou}},\ }\bibfield  {title} {\enquote {\bibinfo
  {title} {Targeted search for the kinematic dipole of the gravitational-wave
  background},}\ }\href {\doibase 10.1103/PhysRevD.106.082005} {\bibfield
  {journal} {\bibinfo  {journal} {Phys. Rev. D}\ }\textbf {\bibinfo {volume}
  {106}},\ \bibinfo {pages} {082005} (\bibinfo {year} {2022})}\BibitemShut
  {NoStop}%
\bibitem [{\citenamefont {Abbott}\ \emph
  {et~al.}(2021{\natexlab{a}})\citenamefont {Abbott}, \citenamefont {Abbott}
  \emph {et~al.}}]{O3directional}%
  \BibitemOpen
  \bibfield  {author} {\bibinfo {author} {\bibfnamefont {R.}~\bibnamefont
  {Abbott}}, \bibinfo {author} {\bibfnamefont {T.~D.}\ \bibnamefont {Abbott}},
  \emph {et~al.} (\bibinfo {collaboration} {LIGO Scientific Collaboration,
  Virgo Collaboration, and KAGRA Collaboration}),\ }\bibfield  {title}
  {\enquote {\bibinfo {title} {Search for anisotropic gravitational-wave
  backgrounds using data from advanced ligo and advanced virgo's first three
  observing runs},}\ }\href {\doibase 10.1103/PhysRevD.104.022005} {\bibfield
  {journal} {\bibinfo  {journal} {Phys. Rev. D}\ }\textbf {\bibinfo {volume}
  {104}},\ \bibinfo {pages} {022005} (\bibinfo {year}
  {2021}{\natexlab{a}})}\BibitemShut {NoStop}%
\bibitem [{\citenamefont {Ballmer}(2006)}]{Ballmer_2006}%
  \BibitemOpen
  \bibfield  {author} {\bibinfo {author} {\bibfnamefont {Stefan~W}\
  \bibnamefont {Ballmer}},\ }\bibfield  {title} {\enquote {\bibinfo {title} {A
  radiometer for stochastic gravitational waves},}\ }\href {\doibase
  10.1088/0264-9381/23/8/s23} {\bibfield  {journal} {\bibinfo  {journal}
  {Classical and Quantum Gravity}\ }\textbf {\bibinfo {volume} {23}},\ \bibinfo
  {pages} {S179--S185} (\bibinfo {year} {2006})}\BibitemShut {NoStop}%
\bibitem [{\citenamefont {Mitra}\ \emph {et~al.}(2008)\citenamefont {Mitra},
  \citenamefont {Dhurandhar}, \citenamefont {Souradeep}, \citenamefont
  {Lazzarini}, \citenamefont {Mandic} \emph {et~al.}}]{mitra07}%
  \BibitemOpen
  \bibfield  {author} {\bibinfo {author} {\bibfnamefont {Sanjit}\ \bibnamefont
  {Mitra}}, \bibinfo {author} {\bibfnamefont {Sanjeev}\ \bibnamefont
  {Dhurandhar}}, \bibinfo {author} {\bibfnamefont {Tarun}\ \bibnamefont
  {Souradeep}}, \bibinfo {author} {\bibfnamefont {Albert}\ \bibnamefont
  {Lazzarini}}, \bibinfo {author} {\bibfnamefont {Vuk}\ \bibnamefont {Mandic}},
   \emph {et~al.},\ }\bibfield  {title} {\enquote {\bibinfo {title}
  {{Gravitational wave radiometry: Mapping a stochastic gravitational wave
  background}},}\ }\href {\doibase 10.1103/PhysRevD.77.042002} {\bibfield
  {journal} {\bibinfo  {journal} {Phys.Rev.}\ }\textbf {\bibinfo {volume}
  {D77}},\ \bibinfo {pages} {042002} (\bibinfo {year} {2008})},\ \Eprint
  {http://arxiv.org/abs/0708.2728} {arXiv:0708.2728 [gr-qc]} \BibitemShut
  {NoStop}%
\bibitem [{\citenamefont {Thrane}\ \emph {et~al.}(2009)\citenamefont {Thrane},
  \citenamefont {Ballmer}, \citenamefont {Romano}, \citenamefont {Mitra},
  \citenamefont {Talukder}, \citenamefont {Bose},\ and\ \citenamefont
  {Mandic}}]{eric_sph}%
  \BibitemOpen
  \bibfield  {author} {\bibinfo {author} {\bibfnamefont {Eric}\ \bibnamefont
  {Thrane}}, \bibinfo {author} {\bibfnamefont {Stefan}\ \bibnamefont
  {Ballmer}}, \bibinfo {author} {\bibfnamefont {Joseph~D.}\ \bibnamefont
  {Romano}}, \bibinfo {author} {\bibfnamefont {Sanjit}\ \bibnamefont {Mitra}},
  \bibinfo {author} {\bibfnamefont {Dipongkar}\ \bibnamefont {Talukder}},
  \bibinfo {author} {\bibfnamefont {Sukanta}\ \bibnamefont {Bose}}, \ and\
  \bibinfo {author} {\bibfnamefont {Vuk}\ \bibnamefont {Mandic}},\ }\bibfield
  {title} {\enquote {\bibinfo {title} {Probing the anisotropies of a stochastic
  gravitational-wave background using a network of ground-based laser
  interferometers},}\ }\href {\doibase 10.1103/PhysRevD.80.122002} {\bibfield
  {journal} {\bibinfo  {journal} {Phys. Rev. D}\ }\textbf {\bibinfo {volume}
  {80}},\ \bibinfo {pages} {122002} (\bibinfo {year} {2009})}\BibitemShut
  {NoStop}%
\bibitem [{\citenamefont {Suresh}\ \emph {et~al.}(2021)\citenamefont {Suresh},
  \citenamefont {Ain},\ and\ \citenamefont {Mitra}}]{pystoch_sph}%
  \BibitemOpen
  \bibfield  {author} {\bibinfo {author} {\bibfnamefont {Jishnu}\ \bibnamefont
  {Suresh}}, \bibinfo {author} {\bibfnamefont {Anirban}\ \bibnamefont {Ain}}, \
  and\ \bibinfo {author} {\bibfnamefont {Sanjit}\ \bibnamefont {Mitra}},\
  }\bibfield  {title} {\enquote {\bibinfo {title} {Unified mapmaking for an
  anisotropic stochastic gravitational wave background},}\ }\href {\doibase
  10.1103/PhysRevD.103.083024} {\bibfield  {journal} {\bibinfo  {journal}
  {Phys. Rev. D}\ }\textbf {\bibinfo {volume} {103}},\ \bibinfo {pages}
  {083024} (\bibinfo {year} {2021})}\BibitemShut {NoStop}%
\bibitem [{\citenamefont {Xiao}\ \emph {et~al.}(2022)\citenamefont {Xiao},
  \citenamefont {Renzini},\ and\ \citenamefont {Weinstein}}]{liting}%
  \BibitemOpen
  \bibfield  {author} {\bibinfo {author} {\bibfnamefont {Liting}\ \bibnamefont
  {Xiao}}, \bibinfo {author} {\bibfnamefont {Arianna~I.}\ \bibnamefont
  {Renzini}}, \ and\ \bibinfo {author} {\bibfnamefont {Alan~J.}\ \bibnamefont
  {Weinstein}},\ }\href {\doibase 10.48550/ARXIV.2211.10010} {\enquote
  {\bibinfo {title} {Model-independent search for anisotropies in stochastic
  gravitational-wave backgrounds and application to ligo-virgo's first three
  observing runs},}\ } (\bibinfo {year} {2022})\BibitemShut {NoStop}%
\bibitem [{\citenamefont {Ain}\ \emph {et~al.}(2015)\citenamefont {Ain},
  \citenamefont {Dalvi},\ and\ \citenamefont {Mitra}}]{folding}%
  \BibitemOpen
  \bibfield  {author} {\bibinfo {author} {\bibfnamefont {Anirban}\ \bibnamefont
  {Ain}}, \bibinfo {author} {\bibfnamefont {Prathamesh}\ \bibnamefont {Dalvi}},
  \ and\ \bibinfo {author} {\bibfnamefont {Sanjit}\ \bibnamefont {Mitra}},\
  }\bibfield  {title} {\enquote {\bibinfo {title} {{Fast Gravitational Wave
  Radiometry using Data Folding}},}\ }\href {\doibase
  10.1103/PhysRevD.92.022003} {\bibfield  {journal} {\bibinfo  {journal} {Phys.
  Rev.}\ }\textbf {\bibinfo {volume} {D92}},\ \bibinfo {pages} {022003}
  (\bibinfo {year} {2015})},\ \Eprint {http://arxiv.org/abs/1504.01714}
  {arXiv:1504.01714 [gr-qc]} \BibitemShut {NoStop}%
\bibitem [{\citenamefont {Ain}\ \emph {et~al.}(2018)\citenamefont {Ain},
  \citenamefont {Suresh},\ and\ \citenamefont {Mitra}}]{pystoch}%
  \BibitemOpen
  \bibfield  {author} {\bibinfo {author} {\bibfnamefont {Anirban}\ \bibnamefont
  {Ain}}, \bibinfo {author} {\bibfnamefont {Jishnu}\ \bibnamefont {Suresh}}, \
  and\ \bibinfo {author} {\bibfnamefont {Sanjit}\ \bibnamefont {Mitra}},\
  }\bibfield  {title} {\enquote {\bibinfo {title} {{Very fast stochastic
  gravitational wave background map making using folded data}},}\ }\href
  {\doibase 10.1103/PhysRevD.98.024001} {\bibfield  {journal} {\bibinfo
  {journal} {Phys. Rev.}\ }\textbf {\bibinfo {volume} {D98}},\ \bibinfo {pages}
  {024001} (\bibinfo {year} {2018})},\ \Eprint
  {http://arxiv.org/abs/1803.08285} {arXiv:1803.08285 [gr-qc]} \BibitemShut
  {NoStop}%
\bibitem [{\citenamefont {Abbott}\ \emph {et~al.}(2022)\citenamefont {Abbott}
  \emph {et~al.}}]{asaf_lvk}%
  \BibitemOpen
  \bibfield  {author} {\bibinfo {author} {\bibfnamefont {R.}~\bibnamefont
  {Abbott}} \emph {et~al.} (\bibinfo {collaboration} {LIGO Scientific
  Collaboration, the Virgo Collaboration, and the KAGRA Collaboration}),\
  }\bibfield  {title} {\enquote {\bibinfo {title} {All-sky, all-frequency
  directional search for persistent gravitational waves from advanced ligo's
  and advanced virgo's first three observing runs},}\ }\href {\doibase
  10.1103/PhysRevD.105.122001} {\bibfield  {journal} {\bibinfo  {journal}
  {Phys. Rev. D}\ }\textbf {\bibinfo {volume} {105}},\ \bibinfo {pages}
  {122001} (\bibinfo {year} {2022})}\BibitemShut {NoStop}%
\bibitem [{\citenamefont {Renzini}\ and\ \citenamefont
  {Contaldi}(2019)}]{Renzini_aniso}%
  \BibitemOpen
  \bibfield  {author} {\bibinfo {author} {\bibfnamefont {A.~I.}\ \bibnamefont
  {Renzini}}\ and\ \bibinfo {author} {\bibfnamefont {C.~R.}\ \bibnamefont
  {Contaldi}},\ }\bibfield  {title} {\enquote {\bibinfo {title} {Improved
  limits on a stochastic gravitational-wave background and its anisotropies
  from advanced ligo o1 and o2 runs},}\ }\href {\doibase
  10.1103/PhysRevD.100.063527} {\bibfield  {journal} {\bibinfo  {journal}
  {Phys. Rev. D}\ }\textbf {\bibinfo {volume} {100}},\ \bibinfo {pages}
  {063527} (\bibinfo {year} {2019})}\BibitemShut {NoStop}%
\bibitem [{\citenamefont {Aasi}\ \emph {et~al.}(2015)\citenamefont {Aasi},
  \citenamefont {Abbott} \emph {et~al.}}]{advLIGO}%
  \BibitemOpen
  \bibfield  {author} {\bibinfo {author} {\bibfnamefont {J}~\bibnamefont
  {Aasi}}, \bibinfo {author} {\bibfnamefont {B~P}\ \bibnamefont {Abbott}},
  \emph {et~al.},\ }\bibfield  {title} {\enquote {\bibinfo {title} {Advanced
  {LIGO}},}\ }\href {\doibase 10.1088/0264-9381/32/7/074001} {\bibfield
  {journal} {\bibinfo  {journal} {Classical and Quantum Gravity}\ }\textbf
  {\bibinfo {volume} {32}},\ \bibinfo {pages} {074001} (\bibinfo {year}
  {2015})}\BibitemShut {NoStop}%
\bibitem [{\citenamefont {Acernese}\ \emph {et~al.}(2014)\citenamefont
  {Acernese}, \citenamefont {Agathos} \emph {et~al.}}]{advVirgo}%
  \BibitemOpen
  \bibfield  {author} {\bibinfo {author} {\bibfnamefont {F}~\bibnamefont
  {Acernese}}, \bibinfo {author} {\bibfnamefont {M}~\bibnamefont {Agathos}},
  \emph {et~al.},\ }\bibfield  {title} {\enquote {\bibinfo {title} {Advanced
  virgo: a second-generation interferometric gravitational wave detector},}\
  }\href {\doibase 10.1088/0264-9381/32/2/024001} {\bibfield  {journal}
  {\bibinfo  {journal} {Classical and Quantum Gravity}\ }\textbf {\bibinfo
  {volume} {32}},\ \bibinfo {pages} {024001} (\bibinfo {year}
  {2014})}\BibitemShut {NoStop}%
\bibitem [{\citenamefont {Romano}\ and\ \citenamefont
  {Cornish}(2017)}]{romanoreview}%
  \BibitemOpen
  \bibfield  {author} {\bibinfo {author} {\bibfnamefont {Joseph~D.}\
  \bibnamefont {Romano}}\ and\ \bibinfo {author} {\bibfnamefont {Neil~J.}\
  \bibnamefont {Cornish}},\ }\bibfield  {title} {\enquote {\bibinfo {title}
  {{Detection methods for stochastic gravitational-wave backgrounds: a unified
  treatment}},}\ }\href {\doibase 10.1007/s41114-017-0004-1} {\bibfield
  {journal} {\bibinfo  {journal} {Living Rev. Rel.}\ }\textbf {\bibinfo
  {volume} {20}},\ \bibinfo {pages} {2} (\bibinfo {year} {2017})},\ \Eprint
  {http://arxiv.org/abs/1608.06889} {arXiv:1608.06889 [gr-qc]} \BibitemShut
  {NoStop}%
\bibitem [{\citenamefont {{Planck Collaboration}}\ \emph
  {et~al.}(2016)\citenamefont {{Planck Collaboration}} \emph
  {et~al.}}]{HubblePlanck}%
  \BibitemOpen
  \bibfield  {author} {\bibinfo {author} {\bibnamefont {{Planck
  Collaboration}}} \emph {et~al.},\ }\bibfield  {title} {\enquote {\bibinfo
  {title} {Planck 2015 results - xiii. cosmological parameters},}\ }\href
  {\doibase 10.1051/0004-6361/201525830} {\bibfield  {journal} {\bibinfo
  {journal} {A\&A}\ }\textbf {\bibinfo {volume} {594}},\ \bibinfo {pages} {A13}
  (\bibinfo {year} {2016})}\BibitemShut {NoStop}%
\bibitem [{\citenamefont {Abbott}\ \emph {et~al.}(2017)\citenamefont {Abbott},
  \citenamefont {Abbott} \emph {et~al.}}]{PhysRevLett.118.121102}%
  \BibitemOpen
  \bibfield  {author} {\bibinfo {author} {\bibfnamefont {B.~P.}\ \bibnamefont
  {Abbott}}, \bibinfo {author} {\bibfnamefont {R.}~\bibnamefont {Abbott}},
  \emph {et~al.} (\bibinfo {collaboration} {LIGO Scientific Collaboration and
  Virgo Collaboration}),\ }\bibfield  {title} {\enquote {\bibinfo {title}
  {Directional limits on persistent gravitational waves from advanced ligo's
  first observing run},}\ }\href {\doibase 10.1103/PhysRevLett.118.121102}
  {\bibfield  {journal} {\bibinfo  {journal} {Phys. Rev. Lett.}\ }\textbf
  {\bibinfo {volume} {118}},\ \bibinfo {pages} {121102} (\bibinfo {year}
  {2017})}\BibitemShut {NoStop}%
\bibitem [{\citenamefont {Floden}\ \emph {et~al.}(2022)\citenamefont {Floden},
  \citenamefont {Mandic}, \citenamefont {Matas},\ and\ \citenamefont
  {Tsukada}}]{Erik_Floden}%
  \BibitemOpen
  \bibfield  {author} {\bibinfo {author} {\bibfnamefont {Erik}\ \bibnamefont
  {Floden}}, \bibinfo {author} {\bibfnamefont {Vuk}\ \bibnamefont {Mandic}},
  \bibinfo {author} {\bibfnamefont {Andrew}\ \bibnamefont {Matas}}, \ and\
  \bibinfo {author} {\bibfnamefont {Leo}\ \bibnamefont {Tsukada}},\ }\bibfield
  {title} {\enquote {\bibinfo {title} {Angular resolution of the search for
  anisotropic stochastic gravitational-wave background with terrestrial
  gravitational-wave detectors},}\ }\href {\doibase
  10.1103/PhysRevD.106.023010} {\bibfield  {journal} {\bibinfo  {journal}
  {Phys. Rev. D}\ }\textbf {\bibinfo {volume} {106}},\ \bibinfo {pages}
  {023010} (\bibinfo {year} {2022})}\BibitemShut {NoStop}%
\bibitem [{\citenamefont {Jenkins}(2022)}]{jenkins_thesis}%
  \BibitemOpen
  \bibfield  {author} {\bibinfo {author} {\bibfnamefont {Alexander~C.}\
  \bibnamefont {Jenkins}},\ }\href@noop {} {\enquote {\bibinfo {title}
  {Cosmology and fundamental physics in the era of gravitational-wave
  astronomy},}\ } (\bibinfo {year} {2022}),\ \Eprint
  {http://arxiv.org/abs/2202.05105} {arXiv:2202.05105 [gr-qc]} \BibitemShut
  {NoStop}%
\bibitem [{\citenamefont {Christensen}(1992)}]{christ92}%
  \BibitemOpen
  \bibfield  {author} {\bibinfo {author} {\bibfnamefont {Nelson}\ \bibnamefont
  {Christensen}},\ }\bibfield  {title} {\enquote {\bibinfo {title} {Measuring
  the stochastic gravitational-radiation background with laser-interferometric
  antennas},}\ }\href {\doibase 10.1103/PhysRevD.46.5250} {\bibfield  {journal}
  {\bibinfo  {journal} {Phys. Rev. D}\ }\textbf {\bibinfo {volume} {46}},\
  \bibinfo {pages} {5250--5266} (\bibinfo {year} {1992})}\BibitemShut {NoStop}%
\bibitem [{\citenamefont {Finn}\ \emph {et~al.}(2009)\citenamefont {Finn},
  \citenamefont {Larson},\ and\ \citenamefont {Romano}}]{ORF_Finn}%
  \BibitemOpen
  \bibfield  {author} {\bibinfo {author} {\bibfnamefont {Lee~Samuel}\
  \bibnamefont {Finn}}, \bibinfo {author} {\bibfnamefont {Shane~L.}\
  \bibnamefont {Larson}}, \ and\ \bibinfo {author} {\bibfnamefont {Joseph~D.}\
  \bibnamefont {Romano}},\ }\bibfield  {title} {\enquote {\bibinfo {title}
  {{Detecting a Stochastic Gravitational-Wave Background: The Overlap Reduction
  Function}},}\ }\href {\doibase 10.1103/PhysRevD.79.062003} {\bibfield
  {journal} {\bibinfo  {journal} {Phys. Rev.}\ }\textbf {\bibinfo {volume}
  {D79}},\ \bibinfo {pages} {062003} (\bibinfo {year} {2009})},\ \Eprint
  {http://arxiv.org/abs/0811.3582} {arXiv:0811.3582 [gr-qc]} \BibitemShut
  {NoStop}%
\bibitem [{\citenamefont {Panda}\ \emph {et~al.}(2019)\citenamefont {Panda},
  \citenamefont {Bhagwat}, \citenamefont {Suresh},\ and\ \citenamefont
  {Mitra}}]{sambit}%
  \BibitemOpen
  \bibfield  {author} {\bibinfo {author} {\bibfnamefont {Sambit}\ \bibnamefont
  {Panda}}, \bibinfo {author} {\bibfnamefont {Swetha}\ \bibnamefont {Bhagwat}},
  \bibinfo {author} {\bibfnamefont {Jishnu}\ \bibnamefont {Suresh}}, \ and\
  \bibinfo {author} {\bibfnamefont {Sanjit}\ \bibnamefont {Mitra}},\ }\bibfield
   {title} {\enquote {\bibinfo {title} {Stochastic gravitational wave
  background mapmaking using regularized deconvolution},}\ }\href {\doibase
  10.1103/PhysRevD.100.043541} {\bibfield  {journal} {\bibinfo  {journal}
  {Phys. Rev. D}\ }\textbf {\bibinfo {volume} {100}},\ \bibinfo {pages}
  {043541} (\bibinfo {year} {2019})}\BibitemShut {NoStop}%
\bibitem [{\citenamefont {Collaboration}\ \emph {et~al.}(2022)\citenamefont
  {Collaboration}, \citenamefont {Collaboration},\ and\ \citenamefont
  {Collaboration}}]{ligo_scientific_collaboration_virgo_coll_2022_6326656}%
  \BibitemOpen
  \bibfield  {author} {\bibinfo {author} {\bibfnamefont {LIGO~Scientific}\
  \bibnamefont {Collaboration}}, \bibinfo {author} {\bibfnamefont {Virgo}\
  \bibnamefont {Collaboration}}, \ and\ \bibinfo {author} {\bibfnamefont
  {KAGRA}\ \bibnamefont {Collaboration}},\ }\href {\doibase
  10.5281/zenodo.6326656} {\enquote {\bibinfo {title} {{Folded data for first
  three observing runs of Advanced LIGO and Advanced Virgo}},}\ } (\bibinfo
  {year} {2022})\BibitemShut {NoStop}%
\bibitem [{\citenamefont {Gorski}\ \emph {et~al.}(2005)\citenamefont {Gorski},
  \citenamefont {Hivon}, \citenamefont {Banday}, \citenamefont {Wandelt},
  \citenamefont {Hansen}, \citenamefont {Reinecke},\ and\ \citenamefont
  {Bartelman}}]{HEALPix}%
  \BibitemOpen
  \bibfield  {author} {\bibinfo {author} {\bibfnamefont {K.~M.}\ \bibnamefont
  {Gorski}}, \bibinfo {author} {\bibfnamefont {Eric}\ \bibnamefont {Hivon}},
  \bibinfo {author} {\bibfnamefont {A.~J.}\ \bibnamefont {Banday}}, \bibinfo
  {author} {\bibfnamefont {B.~D.}\ \bibnamefont {Wandelt}}, \bibinfo {author}
  {\bibfnamefont {F.~K.}\ \bibnamefont {Hansen}}, \bibinfo {author}
  {\bibfnamefont {M.}~\bibnamefont {Reinecke}}, \ and\ \bibinfo {author}
  {\bibfnamefont {M.}~\bibnamefont {Bartelman}},\ }\bibfield  {title} {\enquote
  {\bibinfo {title} {{HEALPix - A Framework for high resolution discretization,
  and fast analysis of data distributed on the sphere}},}\ }\href {\doibase
  10.1086/427976} {\bibfield  {journal} {\bibinfo  {journal} {Astrophys. J.}\
  }\textbf {\bibinfo {volume} {622}},\ \bibinfo {pages} {759--771} (\bibinfo
  {year} {2005})},\ \Eprint {http://arxiv.org/abs/astro-ph/0409513}
  {arXiv:astro-ph/0409513 [astro-ph]} \BibitemShut {NoStop}%
\bibitem [{\citenamefont {Zonca}\ \emph {et~al.}(2019)\citenamefont {Zonca},
  \citenamefont {Singer}, \citenamefont {Lenz}, \citenamefont {Reinecke},
  \citenamefont {Rosset}, \citenamefont {Hivon},\ and\ \citenamefont
  {Gorski}}]{Zonca}%
  \BibitemOpen
  \bibfield  {author} {\bibinfo {author} {\bibfnamefont {Andrea}\ \bibnamefont
  {Zonca}}, \bibinfo {author} {\bibfnamefont {Leo~P.}\ \bibnamefont {Singer}},
  \bibinfo {author} {\bibfnamefont {Daniel}\ \bibnamefont {Lenz}}, \bibinfo
  {author} {\bibfnamefont {Martin}\ \bibnamefont {Reinecke}}, \bibinfo {author}
  {\bibfnamefont {Cyrille}\ \bibnamefont {Rosset}}, \bibinfo {author}
  {\bibfnamefont {Eric}\ \bibnamefont {Hivon}}, \ and\ \bibinfo {author}
  {\bibfnamefont {Krzysztof~M.}\ \bibnamefont {Gorski}},\ }\bibfield  {title}
  {\enquote {\bibinfo {title} {healpy: equal area pixelization and spherical
  harmonics transforms for data on the sphere in python},}\ }\href {\doibase
  10.21105/joss.01298} {\bibfield  {journal} {\bibinfo  {journal} {Journal of
  Open Source Software}\ }\textbf {\bibinfo {volume} {4}},\ \bibinfo {pages}
  {1298} (\bibinfo {year} {2019})}\BibitemShut {NoStop}%
\bibitem [{\citenamefont {Abbott}\ \emph
  {et~al.}(2021{\natexlab{b}})\citenamefont {Abbott}, \citenamefont {Abbott}
  \emph {et~al.}}]{O3iso}%
  \BibitemOpen
  \bibfield  {author} {\bibinfo {author} {\bibfnamefont {R.}~\bibnamefont
  {Abbott}}, \bibinfo {author} {\bibfnamefont {T.~D.}\ \bibnamefont {Abbott}},
  \emph {et~al.} (\bibinfo {collaboration} {LIGO Scientific Collaboration,
  Virgo Collaboration, and KAGRA Collaboration}),\ }\bibfield  {title}
  {\enquote {\bibinfo {title} {Upper limits on the isotropic gravitational-wave
  background from advanced ligo and advanced virgo's third observing run},}\
  }\href {\doibase 10.1103/PhysRevD.104.022004} {\bibfield  {journal} {\bibinfo
   {journal} {Phys. Rev. D}\ }\textbf {\bibinfo {volume} {104}},\ \bibinfo
  {pages} {022004} (\bibinfo {year} {2021}{\natexlab{b}})}\BibitemShut
  {NoStop}%
\bibitem [{\citenamefont {Mitra}\ \emph {et~al.}(2004)\citenamefont {Mitra},
  \citenamefont {Sengupta},\ and\ \citenamefont
  {Souradeep}}]{PhysRevD.70.103002}%
  \BibitemOpen
  \bibfield  {author} {\bibinfo {author} {\bibfnamefont {Sanjit}\ \bibnamefont
  {Mitra}}, \bibinfo {author} {\bibfnamefont {Anand~S.}\ \bibnamefont
  {Sengupta}}, \ and\ \bibinfo {author} {\bibfnamefont {Tarun}\ \bibnamefont
  {Souradeep}},\ }\bibfield  {title} {\enquote {\bibinfo {title} {Cmb power
  spectrum estimation using noncircular beams},}\ }\href {\doibase
  10.1103/PhysRevD.70.103002} {\bibfield  {journal} {\bibinfo  {journal} {Phys.
  Rev. D}\ }\textbf {\bibinfo {volume} {70}},\ \bibinfo {pages} {103002}
  (\bibinfo {year} {2004})}\BibitemShut {NoStop}%
\bibitem [{\citenamefont {Cornish}(2001)}]{Neil_2001}%
  \BibitemOpen
  \bibfield  {author} {\bibinfo {author} {\bibfnamefont {Neil~J}\ \bibnamefont
  {Cornish}},\ }\bibfield  {title} {\enquote {\bibinfo {title} {Mapping the
  gravitational-wave background},}\ }\href {\doibase
  10.1088/0264-9381/18/20/307} {\bibfield  {journal} {\bibinfo  {journal}
  {Classical and Quantum Gravity}\ }\textbf {\bibinfo {volume} {18}},\ \bibinfo
  {pages} {4277} (\bibinfo {year} {2001})}\BibitemShut {NoStop}%
\bibitem [{\citenamefont {Mentasti}\ and\ \citenamefont
  {Peloso}(2021)}]{Mentasti_2021}%
  \BibitemOpen
  \bibfield  {author} {\bibinfo {author} {\bibfnamefont {Giorgio}\ \bibnamefont
  {Mentasti}}\ and\ \bibinfo {author} {\bibfnamefont {Marco}\ \bibnamefont
  {Peloso}},\ }\bibfield  {title} {\enquote {\bibinfo {title} {Et sensitivity
  to the anisotropic stochastic gravitational wave background},}\ }\href
  {\doibase 10.1088/1475-7516/2021/03/080} {\bibfield  {journal} {\bibinfo
  {journal} {Journal of Cosmology and Astroparticle Physics}\ }\textbf
  {\bibinfo {volume} {2021}},\ \bibinfo {pages} {080} (\bibinfo {year}
  {2021})}\BibitemShut {NoStop}%
\bibitem [{\citenamefont {Punturo}\ \emph {et~al.}(2010)\citenamefont
  {Punturo}, \citenamefont {Abernathy} \emph {et~al.}}]{ET}%
  \BibitemOpen
  \bibfield  {author} {\bibinfo {author} {\bibfnamefont {M}~\bibnamefont
  {Punturo}}, \bibinfo {author} {\bibfnamefont {M}~\bibnamefont {Abernathy}},
  \emph {et~al.},\ }\bibfield  {title} {\enquote {\bibinfo {title} {The
  einstein telescope: a third-generation gravitational wave observatory},}\
  }\href {\doibase 10.1088/0264-9381/27/19/194002} {\bibfield  {journal}
  {\bibinfo  {journal} {Classical and Quantum Gravity}\ }\textbf {\bibinfo
  {volume} {27}},\ \bibinfo {pages} {194002} (\bibinfo {year}
  {2010})}\BibitemShut {NoStop}%
\bibitem [{\citenamefont {Agarwal}\ \emph {et~al.}(2022)\citenamefont
  {Agarwal}, \citenamefont {Suresh}, \citenamefont {Mandic}, \citenamefont
  {Matas},\ and\ \citenamefont {Regimbau}}]{targeted_galPlane}%
  \BibitemOpen
  \bibfield  {author} {\bibinfo {author} {\bibfnamefont {Deepali}\ \bibnamefont
  {Agarwal}}, \bibinfo {author} {\bibfnamefont {Jishnu}\ \bibnamefont
  {Suresh}}, \bibinfo {author} {\bibfnamefont {Vuk}\ \bibnamefont {Mandic}},
  \bibinfo {author} {\bibfnamefont {Andrew}\ \bibnamefont {Matas}}, \ and\
  \bibinfo {author} {\bibfnamefont {Tania}\ \bibnamefont {Regimbau}},\
  }\bibfield  {title} {\enquote {\bibinfo {title} {Targeted search for the
  stochastic gravitational-wave background from the galactic millisecond pulsar
  population},}\ }\href {\doibase 10.1103/PhysRevD.106.043019} {\bibfield
  {journal} {\bibinfo  {journal} {Phys. Rev. D}\ }\textbf {\bibinfo {volume}
  {106}},\ \bibinfo {pages} {043019} (\bibinfo {year} {2022})}\BibitemShut
  {NoStop}%
\bibitem [{\citenamefont {Bernardeau}\ \emph {et~al.}(2002)\citenamefont
  {Bernardeau}, \citenamefont {Colombi}, \citenamefont {Gaztañaga},\ and\
  \citenamefont {Scoccimarro}}]{BERNARDEAU20021}%
  \BibitemOpen
  \bibfield  {author} {\bibinfo {author} {\bibfnamefont {F.}~\bibnamefont
  {Bernardeau}}, \bibinfo {author} {\bibfnamefont {S.}~\bibnamefont {Colombi}},
  \bibinfo {author} {\bibfnamefont {E.}~\bibnamefont {Gaztañaga}}, \ and\
  \bibinfo {author} {\bibfnamefont {R.}~\bibnamefont {Scoccimarro}},\
  }\bibfield  {title} {\enquote {\bibinfo {title} {Large-scale structure of the
  universe and cosmological perturbation theory},}\ }\href {\doibase
  https://doi.org/10.1016/S0370-1573(02)00135-7} {\bibfield  {journal}
  {\bibinfo  {journal} {Physics Reports}\ }\textbf {\bibinfo {volume} {367}},\
  \bibinfo {pages} {1--248} (\bibinfo {year} {2002})}\BibitemShut {NoStop}%
\bibitem [{\citenamefont {Jenkins}\ and\ \citenamefont
  {Sakellariadou}(2019)}]{PhysRevD.100.063508}%
  \BibitemOpen
  \bibfield  {author} {\bibinfo {author} {\bibfnamefont {Alexander~C.}\
  \bibnamefont {Jenkins}}\ and\ \bibinfo {author} {\bibfnamefont {Mairi}\
  \bibnamefont {Sakellariadou}},\ }\bibfield  {title} {\enquote {\bibinfo
  {title} {Shot noise in the astrophysical gravitational-wave background},}\
  }\href {\doibase 10.1103/PhysRevD.100.063508} {\bibfield  {journal} {\bibinfo
   {journal} {Phys. Rev. D}\ }\textbf {\bibinfo {volume} {100}},\ \bibinfo
  {pages} {063508} (\bibinfo {year} {2019})}\BibitemShut {NoStop}%
\bibitem [{gwo()}]{gwosc}%
  \BibitemOpen
  \href@noop {} {}\bibinfo {howpublished}
  {\url{gw-openscience.org}}\BibitemShut {NoStop}%
\bibitem [{\citenamefont {van~der Walt}\ \emph {et~al.}(2011)\citenamefont
  {van~der Walt}, \citenamefont {Colbert},\ and\ \citenamefont
  {Varoquaux}}]{5725236}%
  \BibitemOpen
  \bibfield  {author} {\bibinfo {author} {\bibfnamefont {Stefan}\ \bibnamefont
  {van~der Walt}}, \bibinfo {author} {\bibfnamefont {S.~Chris}\ \bibnamefont
  {Colbert}}, \ and\ \bibinfo {author} {\bibfnamefont {Gael}\ \bibnamefont
  {Varoquaux}},\ }\bibfield  {title} {\enquote {\bibinfo {title} {The numpy
  array: A structure for efficient numerical computation},}\ }\href {\doibase
  10.1109/MCSE.2011.37} {\bibfield  {journal} {\bibinfo  {journal} {Computing
  in Science Engineering}\ }\textbf {\bibinfo {volume} {13}},\ \bibinfo {pages}
  {22--30} (\bibinfo {year} {2011})}\BibitemShut {NoStop}%
\bibitem [{\citenamefont {Virtanen}\ \emph {et~al.}(2020)\citenamefont
  {Virtanen} \emph {et~al.}}]{Virtanen2020}%
  \BibitemOpen
  \bibfield  {author} {\bibinfo {author} {\bibfnamefont {Pauli}\ \bibnamefont
  {Virtanen}} \emph {et~al.},\ }\bibfield  {title} {\enquote {\bibinfo {title}
  {Scipy 1.0: fundamental algorithms for scientific computing in python},}\
  }\href {\doibase 10.1038/s41592-019-0686-2} {\bibfield  {journal} {\bibinfo
  {journal} {Nature Methods}\ }\textbf {\bibinfo {volume} {17}},\ \bibinfo
  {pages} {261--272} (\bibinfo {year} {2020})}\BibitemShut {NoStop}%
\bibitem [{\citenamefont {Hunter}(2007)}]{Hunter:2007}%
  \BibitemOpen
  \bibfield  {author} {\bibinfo {author} {\bibfnamefont {J.~D.}\ \bibnamefont
  {Hunter}},\ }\bibfield  {title} {\enquote {\bibinfo {title} {Matplotlib: A 2d
  graphics environment},}\ }\href {\doibase 10.1109/MCSE.2007.55} {\bibfield
  {journal} {\bibinfo  {journal} {Computing in Science \& Engineering}\
  }\textbf {\bibinfo {volume} {9}},\ \bibinfo {pages} {90--95} (\bibinfo {year}
  {2007})}\BibitemShut {NoStop}%
\bibitem [{\citenamefont {Zee}(2003)}]{Zee:2003mt}%
  \BibitemOpen
  \bibfield  {author} {\bibinfo {author} {\bibfnamefont {A.}~\bibnamefont
  {Zee}},\ }\href@noop {} {\emph {\bibinfo {title} {{Quantum field theory in a
  nutshell}}}}\ (\bibinfo  {publisher} {Princeton University Press, 41 William
  Street, Princeton, New Jersey 08540},\ \bibinfo {year} {2003})\BibitemShut
  {NoStop}%
\end{thebibliography}%
\end{document}